\documentclass[manuscript]{emulateapj}
\usepackage{natbib}
\usepackage{times}
\usepackage{amsmath}
\usepackage{color}

\newcommand{\rvec}{\mbox{\boldmath$r$}}

\slugcomment{Accepted for publication in The Astrophysical Journal}

\shorttitle{Wide-Field Hubble Space Telescope Observations of the Globular Cluster System in NGC\,1399}
\shortauthors{Puzia et al.}

\begin{document}
\title{Wide-Field Hubble Space Telescope Observations of the Globular Cluster System in NGC\,1399\footnotemark[$\dagger$]}
\footnotetext[$\dagger$]{Based on observations with the NASA/ESA Hubble Space Telescope obtained at the Space Telescope Science Institute, which is operated by the Association of Universities for Research in Astronomy, Incorporated, under NASA contract NAS5-26555.}

\author{Thomas H. Puzia$^{1,2}$, Maurizio Paolillo$^{3,4,5}$, Paul Goudfrooij$^{6}$, Thomas J. Maccarone$^{7}$, \\ Giuseppina Fabbiano$^{8}$, Lorella Angelini$^{9}$}
\affil{
$^{1}$Institute of Astrophysics, Pontificia Universidad Cat\'{o}lica de Chile, Avenida Vicu\~{n}a Mackenna 4860, Macul, 7820436, Santiago, Chile \\
$^{2}$Herzberg Institute of Astrophysics, 5071 West Saanich Road, Victoria, BC V9E 2E7, Canada \\
$^{3}$Department of Physical Sciences, University of Napoli Federico II, via Cinthia 9, 80126 Napoli, Italy \\
$^{4}$INFN - Napoli Unit, Dept. of Physical Sciences, via Cinthia 9, 80126, Napoli, Italy \\
$^{5}$Agenzia Spaziale Italiana Science Data Center, Via del Politecnico snc, 00133, Roma, Italy \\
$^{6}$Space Telescope Science Institute, 3700 San Martin Drive, Baltimore, MD 21218, USA \\
$^{7}$Texas Tech University, Physics Department, Box 41051, Lubbock, TX 79409, USA \\
$^{8}$Harvard-Smithsonian Center for Astrophysics, 60 Garden Street, Cambridge, MA 02138, USA \\
$^{9}$Laboratory for X-Ray Astrophysics, NASA Goddard Space Flight Center, Greenbelt, MD 20771, USA }
\email{tpuzia@astro.puc.cl}

\begin{abstract}
We present a comprehensive high spatial-resolution imaging study of globular clusters (GCs) in NGC\,1399, the central giant elliptical cD galaxy in the Fornax galaxy cluster, conducted with the {\it Advanced Camera for Surveys} (ACS) aboard the {\it Hubble Space Telescope} (HST).~Using a novel technique to construct drizzled PSF libraries for HST/ACS data, we accurately determine the fidelity of GC structural parameter measurements from detailed artificial star cluster experiments and show the superior robustness of the GC half-light radius, $r_h$, compared with other GC structural parameters, such as King core and tidal radius. The measurement of $r_h$ for the major fraction of the NGC\,1399 GC system reveals a trend of increasing $r_h$ versus galactocentric distance, $R_{\rm gal}$, out to about 10\,kpc and a flat relation beyond. This trend is very similar for blue and red GCs which are found to have a mean size ratio of $r_{\rm h, red}/r_{\rm h, blue}\!=\!0.82\pm0.11$ at all galactocentric radii from the core regions of the galaxy out to $\sim\!40$\,kpc.~This suggests that the size difference between blue and red GCs is due to {\it internal} mechanisms related to the evolution of their constituent stellar populations.~Modeling the mass density profile of NGC\,1399 shows that additional {\it external} dynamical mechanisms are required to limit the GC size in the galaxy halo regions to $r_h\!\approx\!2$ pc. We suggest that this may be realized by an exotic GC orbit distribution function, an extended dark matter halo, and/or tidal stress induced by the increased stochasticity in the dwarf halo substructure at larger galactocentric distances. We compare our results with the GC $r_h$ distribution functions in various galaxies and find that the fraction of extended GCs with $r_h\geq5$\,pc is systematically larger in late-type galaxies compared with GC systems in early-type galaxies.~This is likely due to the dynamically more violent evolution of early-type galaxies.~We match our GC $r_h$ measurements with radial velocity data from the literature and split the resulting sample at the median $r_h$ value into compact and extended GCs. We find that compact GCs show a significantly smaller line-of-sight velocity dispersion, $\langle\sigma_{\rm cmp}\rangle\!=\!225\!\pm\!25$ km s$^{-1}$, than their extended counterparts, $\langle\sigma_{\rm ext}\rangle\!=\!317\!\pm\!21$ km s$^{-1}$.~Considering the weaker statistical correlation in the GC $r_h$-color and the GC $r_h$-$R_{\rm gal}$ relations, the more significant GC size-dynamics relation appears to be astrophysically more relevant and hints at the dominant influence of the GC orbit distribution function on the evolution of GC structural parameters.
\end{abstract}


\section{Introduction}
\subsection{Structural Parameters of Extragalactic GCs}
Wide-field studies of massive galaxies provide important benchmarks for comparisons with glo\-bu\-lar cluster (GC) formation and evolution models as well as GC system assembly in the context of galaxy formation scenarios, not only because they define homogeneous and uniform datasets but also due to their simultaneous sampling of galaxy core and halo regions where various different physical processes affect the GC formation and survivability.~In general, GC formation is influenced by small-scale physics that governs star-formation and feedback processes \citep[e.g.][]{murray92, harris94, ee97, hartwick09, murray09} while stellar feedback as well as internal and external dynamical mechanisms determine their early evolution \citep{gieles08, bastianetal08, fall09, chandar09, elmegreen10, mapelli13} and the latter, ultimately, their fate \citep[e.g.][]{gnedin97, vesperini97, vesperini03, chandar10, bekki10}.~The vast dynamical parameter ranges that need to be probed to study the complex interplay of these processes with numerical simulations are still very challenging for today's computers \citep[e.g.][]{kravtsov05, li05, bournaud08, griffen10, schulman12, greif12}.~One simple approach to understand the influence of some of these processes on GC formation and evolution is the empirical study of GC structural parameters and their variation as a function of galactocentric distance.

Detailed GC structural parameters, such as core, half-light, and tidal radius, as well as central surface brightness, concentration, ellipticity, etc.~were, until the past decade, only accessible within the Local Group (LG) due to the limited spatial resolution of ground-based instrumentation \citep[e.g.][]{king68, ill76, kontizas82, elson85, elson88, elson91, elson92, crampton85, demers90, trager95}.~The launch of the {\it Hubble Space Telescope} (HST) catapulted this field to a whole new stratum, making vast numbers of GC systems accessible to high spatial resolution studies \citep{harris13}. In fact, HST still provides our only access to high spatial resolution observations at optical wavelengths.~Several pioneering HST works quickly reached out with their GC half-light radius measurements beyond the LG as far as the Fornax galaxy cluster at $\sim\!20$ Mpc distance \citep[e.g.][]{elson94, fusipecci94, kundu98, kundu99, puzia99, puzia00, zepf99}.~Numerous subsequent studies have used the superior spatial resolution of HST and the relatively large field of view ($\sim\!202\arcsec \!\times\! 202\arcsec$) of the {\it Advanced Camera for Surveys} (ACS) to collect large imaging datasets of extragalactic GC systems, the most homogeneous of which was obtained by the ACS Virgo and Fornax Cluster Surveys (ACSVCS and ACSFCS, see \citealt{cote04} and \citealt{jordan07}, respectively).~These observations set the baseline for systematic studies of GC structural parameters in the central regions of early-type cluster galaxies.~There was quickly mounting consensus among the early HST investigations that the observed central GCs had a rather broad half-light radius distribution with a peak somewhere in the range $\sim\!2\!-\!3$ pc, which led to the suggestion that this peak value may be used as a geometric distance indicator \citep[e.g.][]{kundu01, jordan05}.~Another important finding was that the blue GCs are on average larger than the red GCs.~In particular, within the central regions of galaxies typically observed with HST, blue GCs show $\sim\!20\%$ larger mean half-light radii compared to the red GC sub-population \citep[e.g.][]{kundu98, kundu01, kundu99, puzia99, puzia00, zepf99, larsen01, jordan05, spitler06, harris06, harris09, harris10, blom12, goudfrooij12}.

However, one major limitation of most previous HST studies, targeting extragalactic GC systems such as the ACS Virgo and Fornax cluster surveys, was their limited field of view, using only one HST/ACS pointing per galaxy.~Because of this, these HST studies focused on the core regions of elliptical galaxies covering the inner few kpc (i.e.~$\lesssim 1 R_{\rm eff}$).~The outer parts of rich GC systems in central cluster galaxies were so far missed and mainly observed with ground-based instrumentation at much lower spatial resolutions \citep[e.g.][]{rhode01, rhode04, rhode07}.~The only other ground-based study featuring a wide field of view {\it and} high spatial-resolution was performed by \cite{gomez07} using Magellan/IMACS under exceptional $\sim\!0.5\arcsec$ average seeing conditions to measure half-light radii of 364 radial-velocity confirmed GCs in NGC~5128 (S0/E) out to $\sim\!8 R_{\rm eff}$ of the spheroid light and found no significant correlation between GC half-light radius and projected galactocentric distance, i.e.~$r_h\!\propto\!R^0$, outside $\sim\!1 R_{\rm eff}$.~However, \citeauthor{gomez07} reported that at $\lesssim\!1 R_{\rm eff}$ the red GCs show a steeper $r_h\!-\!R$ relation and on average 30\% smaller sizes than blue GCs.

Other studies using more than single-pointing HST observations conducted GC half-light radius measurements in NGC\,4594 (Sa) out to about $6\,R_{\rm eff}$ of the bulge component \citep[658 GC candidates]{spitler06, harris10}, in NGC\,4365 (E) out to $\sim\!2.4 R_{\rm eff}$ of the spheroid light \citep[659 GC candidates]{blom12}, and in 6 giant elliptical galaxies out to $\sim\!4\!-\!5 R_{\rm eff}$ of their spheroids \citep[altogether 3330 GC candidates]{harris09}.~In the case of NGC\,4594, the inner red GCs are $\sim\!17\%$ smaller than the blue ones, but because of a steeper size-radius relation of the red GC sub-population, this difference becomes insignificant at galactocentric radii $\gtrsim\!2.7 R_{\rm eff}$ of the bulge light.~NGC\,4365 hosts on average $\sim\!32\%$ larger blue GCs compared to their red counterparts and shows a steep size-radius relation, $r_h\!\propto\!R^{(0.49\pm0.04)}$, for the entire GC sample, similar to Milky Way's GC system \citep[][]{vdB91}.~However, \citeauthor{blom12}~do not investigate whether this relation differs between GC sub-populations as a function of projected galactocentric radius.~The composite GC system of the six giant ellipticals studied by \citeauthor{harris09} exhibits a mild relation of the form $r_h\!\propto\!R^{0.11}$ and a $\sim\!17\%$ size difference between red and blue GCs that is independent of projected galactocentric radius.

\subsection{Astrophysical Implications}
In general, the finding of a size difference between blue and red GCs has important astrophysical implications for the understanding of the formation and evolution of GCs and for the usefulness of the peak value of the GC size distribution as geometric distance indicator. Several studies, such as \cite{larsen03}, \cite{jordan04}, and \cite{harris09} put forward models to explain the size difference between blue and red GCs.~Inspired by the Milky Way GC system where a shallow relation exists between GC half-light radius and the 3-dimensional galactocentric distance, $r_h\!\propto\!R_{\rm 3D}^{0.5}$ \citep{vdB91}, \citeauthor{larsen03} suggested that the GC size difference between red and blue GCs in massive ellipticals could be due to the difference in their spatial distribution functions.~Typically, the red GC sub-population would be more centrally concentrated than their blue counterpart and therefore on average smaller, being tidally more truncated by the stronger host galaxy potential.~However, \cite{webb12b} have shown in detailed numerical simulations that the observed GC size difference is unlikely due to projection effects alone.~In contrast to this {\it external effect}, two alternative {\it internal effects} were put forward. Firstly, \cite{jordan04} suggested that the combined effect of mass segregation and shorter stellar lifetimes of more metal-rich stars at a given mass may explain the GC size difference.~This was strictly valid under the assumption that the GC {\it half-mass} radius distribution would be independent of metallicity and that metal-poor and metal-rich GCs were of the same age, which may be at odds with observations \citep[e.g.][]{puzia02, marinfranch09, goudfrooij12}.~This scenario was further developed by \cite{sippel12} and \cite{schulman12} in direct-integration N-body simulations of young, low-mass clusters with and without initial mass segregation, the absence of which was found to be enhancing the GC size difference.~\cite{downing12} performed Monte-Carlo N-body simulations of massive star clusters and found that significant numbers of massive stellar remnants, i.e.~single and binary black holes would boost this GC size difference.~Secondly, \cite{harris09} suggested that more metal-rich proto-GC clouds could cool more efficiently and therefore collapse into a more concentrated quasi-equilibrium state before forming stars than clusters formed from low-metallicity gas.~Any of these three scenarios comes with limiting assumptions and is likely not the single cause for the measured GC size difference as the variety of results described above indicates.

To provide a larger and statistically robust dataset to constrain GC sizes as a function of galactocentric radius, we embarked on a wide-field observing campaign covering a large area with an HST/ACS mosaic out to several effective radii ($>\!5 R_{\rm eff}$) of the diffuse spheroid light around NGC\,1399, the central galaxy in the Fornax galaxy cluster that hosts one of the richest ($\ga6000$ GCs; Specific frequency\footnote{The specific frequency of a GC system is defined as twice the number of GCs brighter than the turn-over luminosity of the GC luminosity function, given by $N_{\rm GC}$, relative to the absolute $V$-band luminosity of the host galaxy, $M_V$, which is normalized to $-15$ mag.~This quantity is defined as the specific frequency of a GC system $S_N\!=\!N_{\rm GC}10^{0.4(M_V+15)}$; see also \cite{georgiev10} and \cite{harris13} for other GC system scaling relations.} $S_N\approx5$) and most extended GC systems in the nearby Universe \citep{dirsch03, faifer04, bassino06}. A significant part of the outer-halo GC system of NGC\,1399 is located hundreds of kpc away from its host and is probing the transition regime between galaxy and cluster potential \citep{ferguson89}.~At the same time, the formation efficiencies of these outer-halo, blue GCs appear to be higher than those of the inner red GCs, $S_N(\mbox{red})\!\approx\!3$ while $S_N(\mbox{blue})\!\approx\!14$ \citep{forte05}. Spectroscopic radial-velocity studies of hundreds of GCs established a very complex multi-component system with the blue GCs being kinematically distinct from the red GC sub-population the latter of which shows dynamics similar to that of the host galaxy diffuse stellar component.~The blue GCs, on the other hand, seem to have been partly accreted from satellite galaxies \citep{schuberth10}.~It is this large auxiliary kinematic dataset that makes the GC system of NGC\,1399 an ideal target for a wide-field, high spatial-resolution study with HST/ACS \citep[in comparison to M87, e.g.][]{epeng09, madrid09} as several hundreds of member stellar systems are robustly separated from the fore- and background in radial velocity space.

In our previous works, we used the dataset from this paper to study the Low Mass X-ray Binary (LMXB) population and the correlation of their properties with GC structural parameters \citep{paolillo11, dago13}, as well as the GC selection techniques based on neural algorithms \citep{brescia12}. Here we focus on the properties of the GC system itself.~Our present paper is organized as follows: in \S2 we present the HST/ACS observations and discuss the details of sub-pixel dithering, \S3 includes a description of the preliminary photometry that enters our structural parameter fitting code, which is introduced and thoroughly tested in \S4.~We present our results in \S5, where we show the large-scale variations of GC structural parameters within NGC\,1399.~We discuss the implications in \S6 and conclude this work in \S7.

\section{Observations}
\begin{figure}[!t]
\centering
\includegraphics[width=8.7cm]{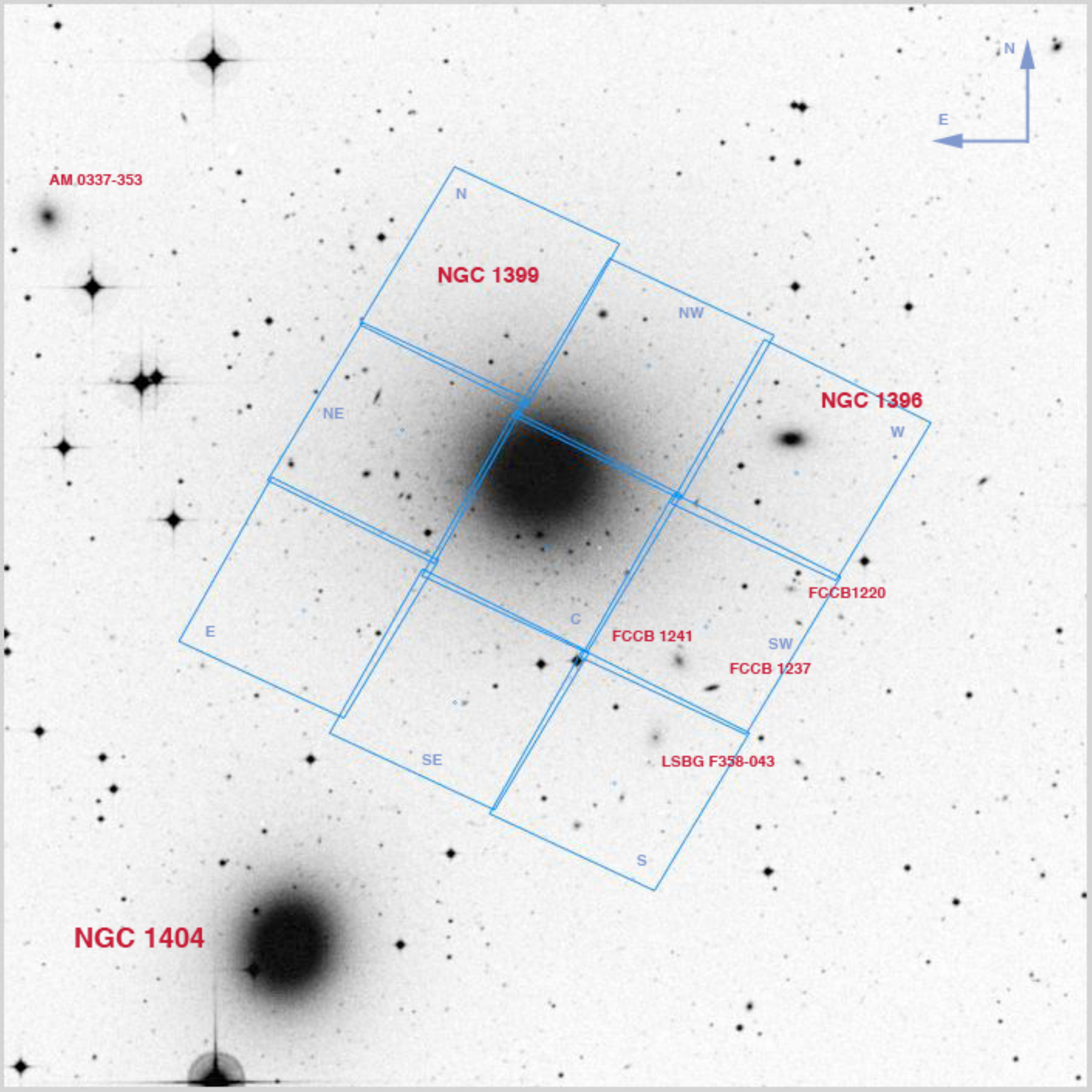}
\caption{Illustration of the 3x3 mosaic of our ACS observations overplotted on a DSS-2 image. Individual tiles and the main galaxies in the field are labeled. The orientation of the image, which measures $20\arcmin\times 20\arcmin$, is indicated in the upper right corner.}
\label{fov}
\end{figure}

\subsection{Field Coverage and Orientation}
All observations were taken as part of the program GO-10129 (PI:Puzia) with the {\it Advanced Camera for Surveys} \citep[ACS;][]{ford03} onboard the {\it Hubble Space Telescope} (HST) in November 2004 and April 2005.~The pointings were arranged in a $3\!\times\!3$ ACS mosaic with a few arcseconds overlap between the individual tiles as illustrated in Figure~\ref{fov}.~To maximize common-field coverage with other imaging and spectroscopy observations (i.e.~{\it Chandra} X-ray imaging, see \citealt{paolillo11}, and VLT ground-based spectroscopy) the entire mosaic is rotated with a position angle of about $-30\,^{\rm o}$ with respect to the meridian and centered on the coordinates: RA (J2000)~$\!=\!03^{\rm h} 38^{\rm m} 28.62^{\rm s}$ and Dec (J2000)~$\!=\!-35^{\rm o} 28\arcmin\ \! 18.9\arcsec$.~Due to scheduling constraints the north, north-east, and north-west tiles were observed with a position angle $-30.467\,^{\rm o}$, while the other six tiles were taken at a position angle $149.552\,^{\rm o}$.~The full mosaic covers roughly $10\arcmin\!\times\!10\arcmin$ arcminutes and extends out to a maximum projected galactocentric distance of 8.76\arcmin\ or $51.3\pm1.0$ kpc with respect to NGC\,1399 \citep[adopting the distance $D\!=\!20.13\!\pm\! 0.4$ Mpc, see][also \citealt{blakeslee09}]{dunn06}. This corresponds to a projected coverage of $\sim\!5.2$ effective radii of the NGC\,1399 diffuse galaxy light \citep{RC3} and $\sim\!4.9$ core radii of the globular cluster system density profile\footnote{\cite{schuberth10} approximate the radial GC system number density distribution with a cored power-law profile of the form $N(R)\!\propto\!((R/R_0)^2\!+\!1)^{-\alpha}$, where the core radius is $R_0\!=\!1.74\arcmin\pm0.27\arcmin$ and the power-law exponent $\alpha=0.84\pm0.02$.} \citep{schuberth10}.

Our filter choice considerations included the optimization of throughput, detector sensitivity, high spatial resolution, and a well-defined transformation to a standard photometric system.~The filter that optimally balances these effects is F606W and was used for all our exposures.~The ACS {\it Wide-Field Channel} (WFC) spatial sampling of the point-spread function (PSF) is sub-critical at the wavelength of our observations (F606W~$\approx4600\!-\!7200$\AA).~If not accounted for, this would introduce aliasing artifacts and significantly degrade the spatial information in the final images, thus hampering the measurement of globular cluster structural parameters at the distance of Fornax.~Each tile was, therefore, observed in a single orbit in four dithered sub-exposures of 527 seconds to allow sub-pixel resampling (see below), yielding a total integration time of 2108 seconds.

\subsection{Data Reduction and Image Combination}
\label{ln:drizz}
The basic data reduction of each ACS/WFC dither set was performed by the ACS data pipeline CALACS \citep{hack03}.~The reduction steps included subtraction of masterbias and masterdark images, correction for flat-field and gain variations, as well as elimination of bad pixels.

\begin{deluxetable}{rl}[!t]
\centering
\tabletypesize{\scriptsize}
\tablecaption{Parameters of the utilized dither pattern\label{tbl-dither}}
\tablewidth{0pt}
\tablehead{\colhead{Parameter} & \colhead{value}}
\startdata
Pattern type                  & ACS-WFC-DITHER-BOX \\
Primary pattern shape  & PARALLELOGRAM \\
Pattern purpose            & DITHER \\
Number of points          & 4 \\
Point spacing                & 0.285\arcsec \\
Line spacing                 & 0.285\arcsec \\
Coordinate frame         & POS-TARG \\
Pattern orient               & 30.155 deg \\
Angle between sides    & 145.82 deg \\
Center pattern              & NO 
\enddata
\end{deluxetable}

For the dithered observations we adopted a slightly modified Hubble Ultra-Deep Field dither pattern, for which the dither parameters are provided for reference in Table~\ref{tbl-dither}. Note that this dither pattern is not designed to cross the ACS inter-chip gap, but to maximize the sub-pixel shift integrity over the full ACS/WFC field of view. In its shape it follows the UDF dither pattern with a 67\% larger step size. 

Each set of four dithered frames was combined into a single image using the {\sc MultiDrizzle} routine v.2.7.0 \citep{koekemoer02}. The software takes care of correcting the geometric field distortions which affect individual ACS exposures and projects all dithered images onto a common grid in which the rectified frames are averaged.~The averaged image is then "blotted" back into each distorted frame to identify and clean cosmic rays and bad pixels/columns by means of comparison of input vs.~averaged image \citep[see][]{fruchter02}. No background subtraction was performed at this stage of data processing. The main background contribution in our fields is due to the NGC 1399 diffuse light and is correctly accounted for in the following structural profile analysis (see Sect.~\ref{ln:profilefit}).

\begin{figure*}[!t]
\centering
\includegraphics[width=17.5cm]{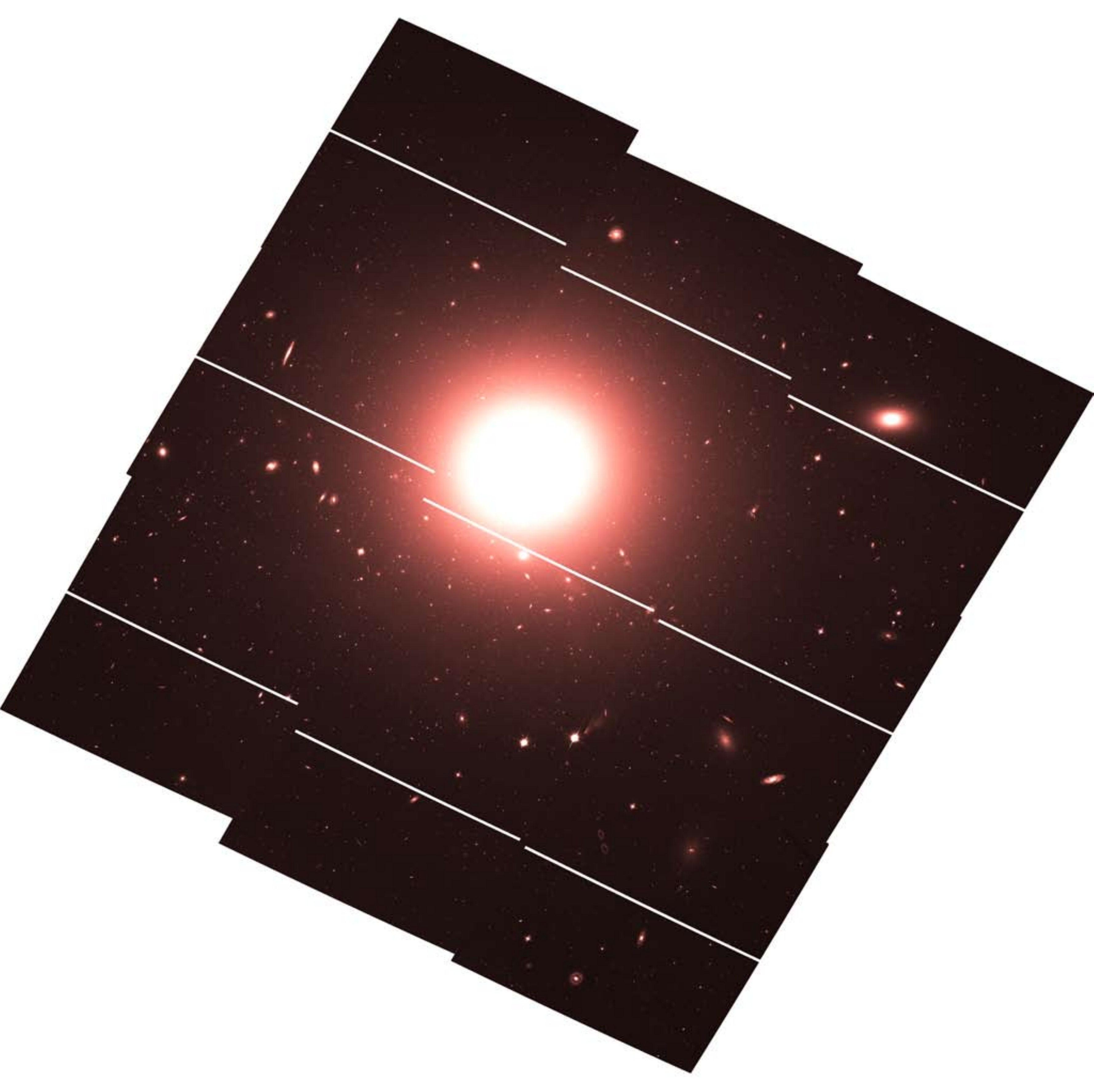}
\caption{Composite field of view of the combined $3\!\times\!3$ ACS mosaic roughly centered on NGC\,1399. The effective area of the observed sky is 99.053 arcmin$^2$ and it covers $\sim\!4.9$ core radii of the globular cluster system in NGC\,1399 \citep{schuberth10} and $\sim\!5.7$ effective radii of the NGC\,1399 diffuse galaxy light \citep{RC3}. The dimensions of this field are 13.78\arcmin\ in RA and 13.75\arcmin\ in Dec. The white lines are the ACS/WFC inter-chip gaps which were not covered by our dither pattern. North is up, east is to the left.}
\label{pdf:full}
\end{figure*}

Similar to the GOODS and UDF datasets, we use the Gaussian drizzle kernel and set the pixel scale to 0.03\arcsec/pix on the final combined images.~This provides a super-Nyquist sampling of the PSF with a FWHM of $\sim\!0.08\arcsec$ at 6000 \AA\ \citep[see also][]{beckwith06}. \cite{rhodes07} find that this combination of Gaussian drizzle kernel and 0.03\arcsec/pix output pixel scale gives minimal aliasing in the final images. \cite{jee07} argue that a Lanczos drizzle kernel with a 0.05\arcsec/pix output scale reduces the PSF width by $\sim\!3\%$ compared to the Gaussian kernel, at the expense that the Lanczos kernel introduces ``cosmetic artifacts in the regions where flux gradients change abruptly" \citep{jee07}.~Since most of our target GCs are likely to have structural parameters at the resolution limit of HST we are expecting strong varying profile gradients for the most compact objects.~We find that noise correlation between neighbouring pixels produces moir\'{e} patterns in the vicinity of bright objects and strong gradients \cite[see also][]{rhodes07}, but this affects only a few blended sources in our dataset. After these considerations and careful visual inspection of the drizzled images we therefore decide to use the Gaussian drizzle kernel with {\sc pixfrac}=0.8 in the subsequent analysis. The combined field is illustrated in Figure~\ref{pdf:full} and has an effective field of view of 99.053 arcmin$^2$.

\begin{figure}[!t]
\centering
\includegraphics[bb=-20 -20 700 700, width=8.5cm]{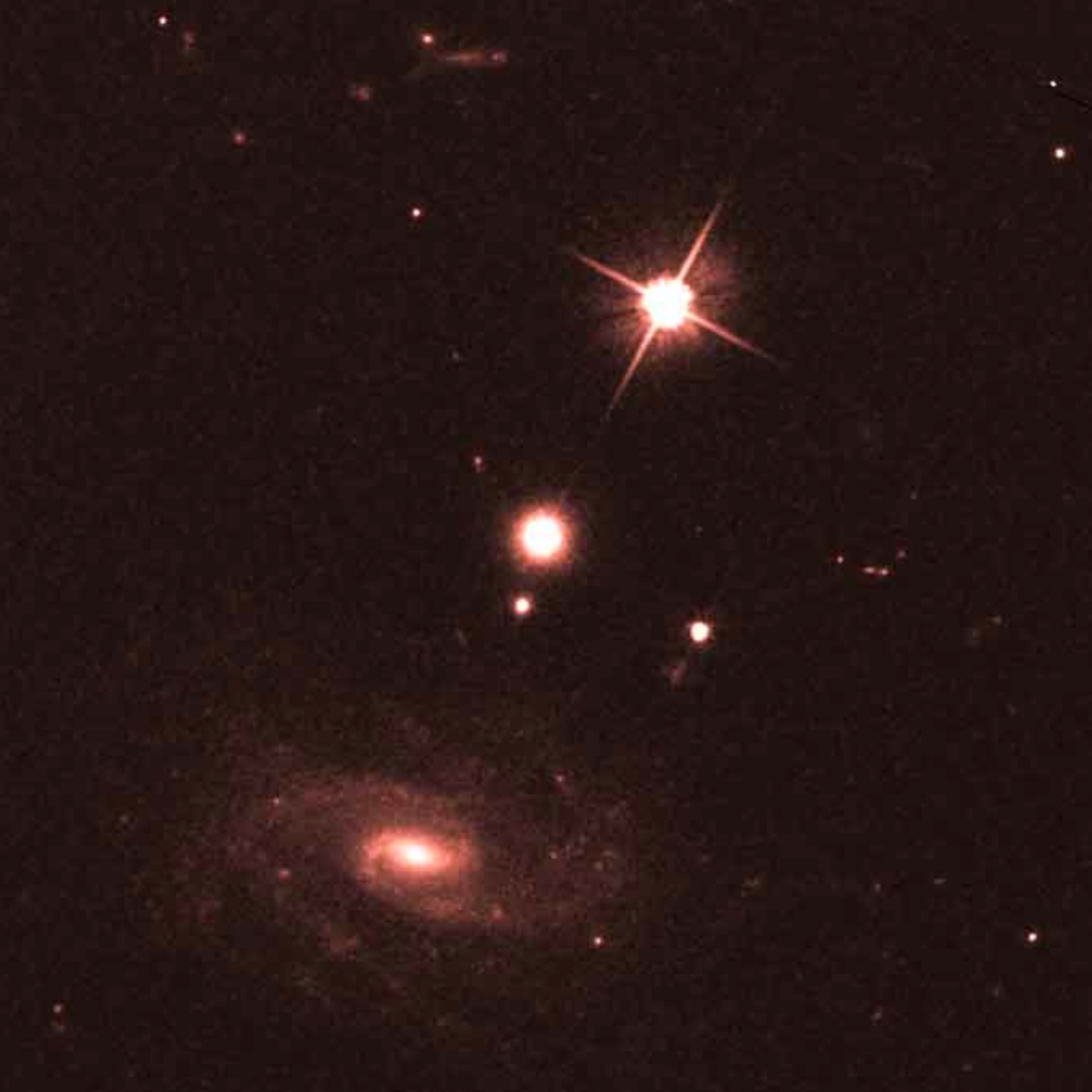}
\caption{This approximately $35\arcsec\!\times\!35\arcsec$ image cutout is an illustration of the data quality of our drizzled HST/ACS mosaic frames, featuring a representative region with a bright foreground star, several resolved compact stellar systems in NGC\,1399, a background spiral galaxy with its own disk star cluster system, and many faint background sources.~North is up, east is to the left.}
\label{pdf:example}
\end{figure}

Using the {\sc MultiDrizzle} software, we also produce weight and error maps representing the final error budget for each pixel, which account for all uncertainties in the reduction process, including bias, flatfield, drizzling, and aliasing effects.~These weight and error maps enter the photometry and structural parameter analysis.~We note that the high spatial resolution of our drizzled images safeguards them from crowding effects, even in the central regions of NGC\,1399, and reveals in every pointing a wealth of detail in object morphology as illustrated in Figure~\ref{pdf:example}.

\section{The Photometric Input Catalog}
\label{ln:photin}
\subsection{Aperture Photometry and Astrometry}
To obtain a rough estimate of the total magnitudes of all detected sources we perform aperture photometry with the {\sc SExtractor} package \citep{bertin96} and measure instrumental magnitudes in apertures of successively growing diameter, i.e.~photometric growth curve analysis.~We use the asymptotic limit of these curves to compute mean photometric corrections from finite aperture sizes to ``infinity''.~Our tests show that an aperture with $0.24\arcsec$ radius maximizes the signal-to-noise ratio (S/N) of the final photometry. Leaving out saturated objects and spurious detections we obtain a mean aperture correction for this optimal aperture size to an "infinite" aperture radius of $\langle\Delta_{\rm F606W}\rangle=-0.14$ mag (with a standard deviation $\sigma\!=\!0.22$ mag).~We also measure the mean photometric correction from the standard 0.5\arcsec\ aperture radius to "infinity" $\langle\Delta_{\rm F606W}\rangle=-0.07$ mag (with a standard deviation $\sigma\!=\!0.14$ mag), which compares well with the suggested value from \cite{sirianni05} of $\langle\Delta_{\rm F606W}\rangle=-0.088$ mag.

We follow the prescriptions of \cite{sirianni05} to calibrate our F606W "infinite"-aperture magnitudes $m_i$ to the broadband $V$-filter in the VEGAMAG filter system.~We include second-order color terms from the synthetic model of \citeauthor{sirianni05} that are applicable for the color range $V\!-\!I\!>\!0.4$ mag and obtain the final photometric calibration equation
\begin{equation}
\label{photcalib}
V_{\rm F606W} = m_i + 26.331 + 0.340\, (V\!-\!I) - 0.038\, (V\!-\!I)^2,
\end{equation}
where we assume a mean $V\!-\!I\!=\!0.95\!\pm\!0.1$ mag for our globular cluster candidates \citep[see e.g.][]{peng06}.~All frames have a minimum background flux level of $\sim\!40\, e^-$ per sub-integration which would correspond to a CTE correction of the order $\la 0.02$ mag across each WFC chip \citep{riess04, vera07}.~Since the average background level in all ACS mosaic tiles is higher than the minimum background flux we do not correct for this negligible photometric offset.~The Galactic foreground extinction in the direction of NGC\,1399 is $E(B\!-\!V)\!=\!0.013$ mag \citep{schlegel98, schlafly11}, which translates into $A_{\rm F606W}\!=\!0.038$ mag using the \cite{schlegel98} reddening curve. The total uncertainty of the photometric calibration in Equation~\ref{photcalib} formally amounts to $\sim\!0.089$ mag. However, when we consider the small CTE corrections, a color mismatch of $\sim0.1$ mag in Equation~\ref{photcalib} for GCs with extreme $V\!-\!I$ colors, and potential differential reddening of $\sim\!0.05$ mag across the ACS mosaic field, we estimate that our final photometry is accurate to $\Delta V_{\rm F606W}\approx 0.1$ mag.~It is important to note that at this point we are not concerned with achieving photometry of the highest possible quality but providing first-guess input catalogs for our profile fitting routine.

To be able to match source detections taken with other telescopes we compute an absolute astrometric solution for each of the nine ACS tiles.~We select 40 bright unsaturated stars distributed homogeneously over the entire mosaic and match their positions with those of stars from the USNO-B1 catalog\footnote{\url{http://tdc-www.harvard.edu/software/catalogs/ub1.html}} \citep{monet03} to obtain the world coordinate solution (WCS) for each tile. The final WCS accuracy across the entire mosaic is $\sim0.2\arcsec$.

\subsection{Object Classification}
\label{ln:gcc}

In the following we describe the object detection and classification schemes that were used to define a photometrically-selected globular cluster candidate (GCC) sample for which we later measure structural parameters (see Sect.~\ref{ln:sp}).~On the drizzled stack images we measure object coordinates, the background level, Kron radius\footnote{The Kron radius is defined as $r_k=\sum rI(r)/\sum I(r)$. A circular aperture of radius $2r_k$ encloses $\geq90$\% of an object's flux independent of its magnitude \citep{kron80}.}, isophotal area, FWHM, ellipticity, position angle, and the {\sc SExtractor} quality flag parameter of each detection that had at least 20 pixels approximately $1.6\,\sigma$ above the background noise, corresponding to S/N~$\approx 6$.~The error images, produced during the drizzle procedure, were used as weight maps in the detection process to account for the varying NGC\,1399 surface brightness.

\begin{figure*}[!th]
\centering
\includegraphics[width=6.5cm]{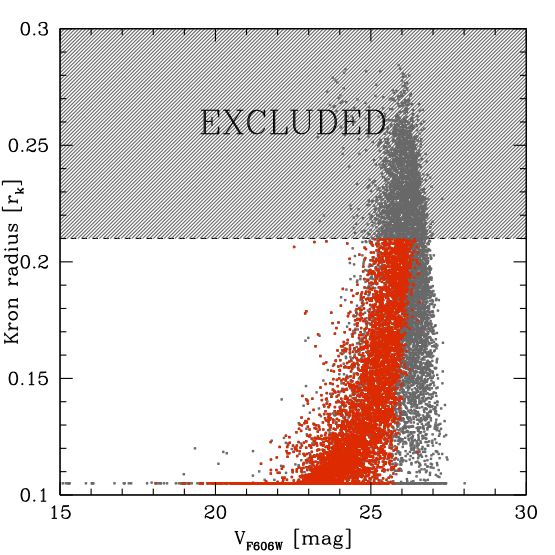}
\includegraphics[width=6.5cm]{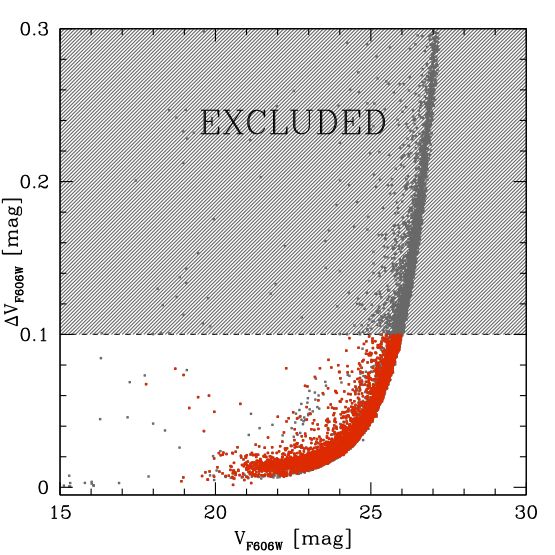}
\includegraphics[width=6.5cm]{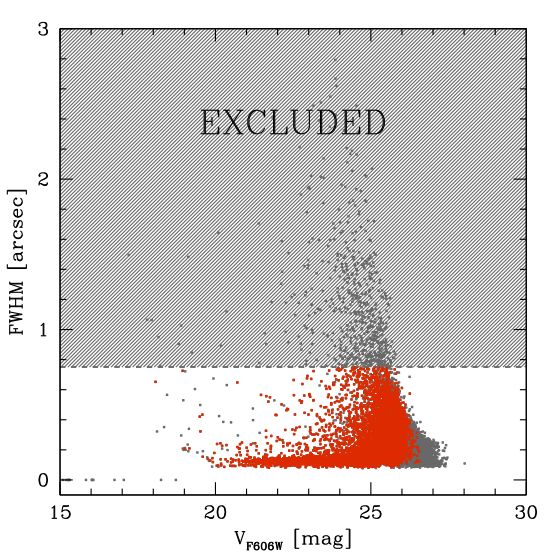}
\includegraphics[width=6.5cm]{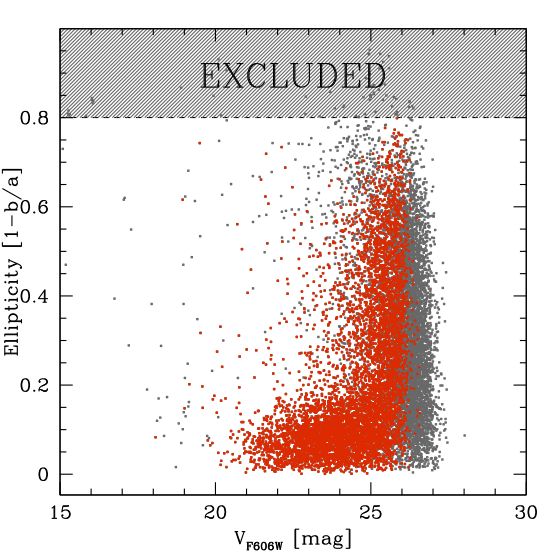}
\caption{The above panels show the photometric diagnostics (from the upper right panel counter-clockwise) photometric uncertainty, Kron radius, FWHM, and ellipticity as a function of $V_{\rm F606W}$ magnitude for all detected objects in the ACS mosaic ({\it grey data}). The hatched regions indicate excluded objects excluded by the selection cuts that are used to reject extended and/or amorphous background objects and image artifacts. The selected objects ({\it red data}) are used as input sample to measure their structural parameters. See Section~\ref{ln:gcc} for details.}
\label{gcsel}
\end{figure*}

\begin{figure}[!t]
\centering
\includegraphics[width=8.9cm]{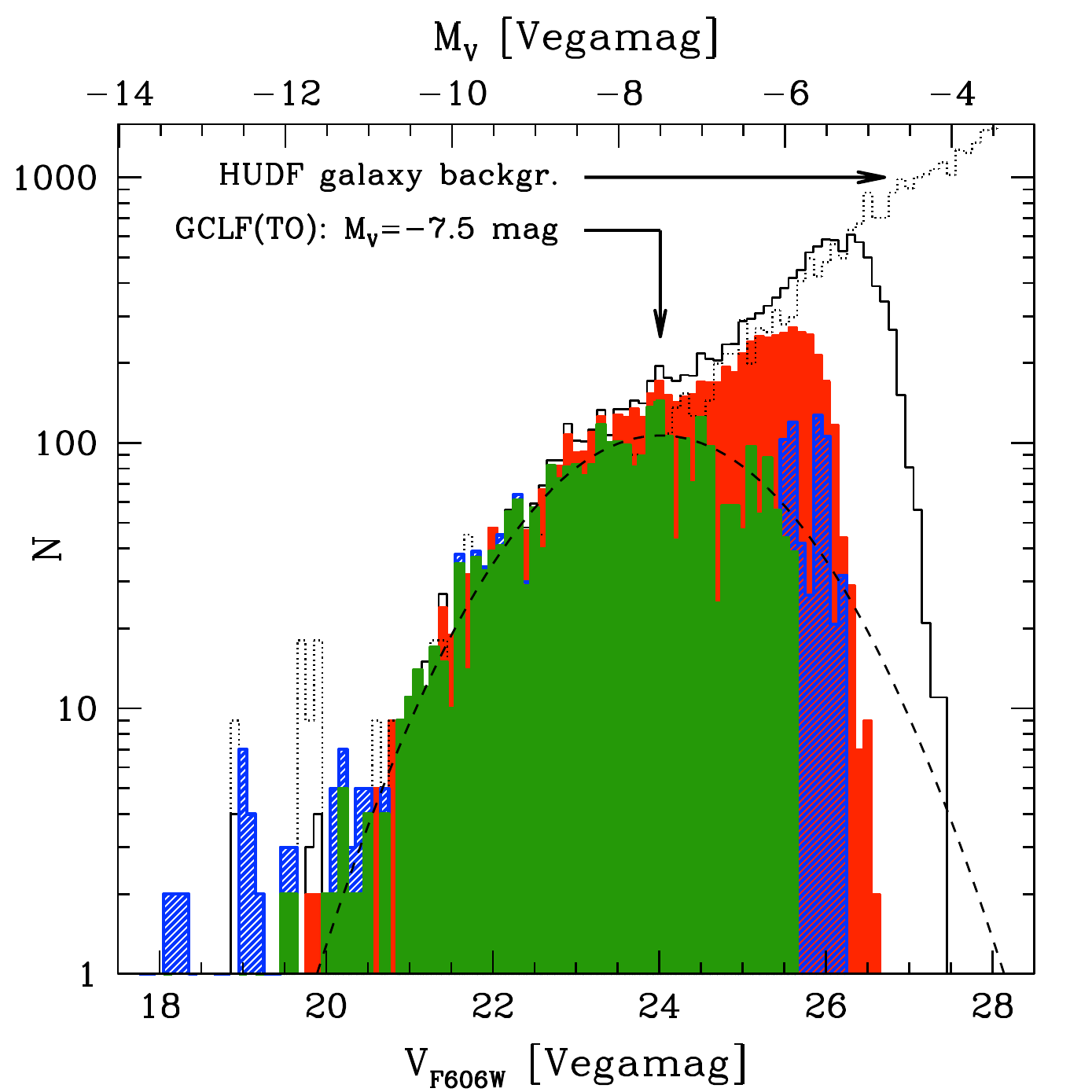}
\caption{Luminosity distribution of all F606W detections around NGC\,1399 in our $3\!\times\!3$ ACS mosaic (solid open histogram) and the background galaxy contribution estimated from the HUDF based on the same object detection criteria (dotted open histogram). Our pre-selected sample that enters the structural parameter measuring routine is shown as solid red histogram. To illustrate the influence of the background galaxy population on the faint end of our input catalog sample we statistically subtract the background galaxy distribution (dotted open histogram) from our initial photometric sample (solid open histogram) and show the result as the hatched blue histogram. The corresponding result after filtering with the photometric pre-selection (see Sect.~\ref{ln:gcc} for details) is shown as solid green histogram. This is in remarkable agreement with the expected classic GC luminosity function with $M_V({\rm TO})\!=\!-7.5$ mag and $\sigma_{\rm GCLF}\!=\!1.4$ mag, which is indicated as dotted curve \citep[e.g.][]{richtler03}.~The top axis shows absolute magnitudes assuming a distance of $D\!=\!20.13$ Mpc.}
\label{gclf}
\end{figure}

Rather than trying to find the optimal source parameters to select high-probability GCCs, we adjust our classification parameters to reject clearly extended and/or amorphous background objects and image artifacts.~Visual inspection of the individual frames shows that a very reliable rejection of clearly extended background sources and image artifacts is provided by the following parameter cuts: $\Delta V_{\rm F606W}\!<\!0.1$ mag, Kron radius $r_k<0.21$\arcsec, FWHM~$<0.75$\arcsec, ellipticity ($1-b/a)<0.8$ (see Figure~\ref{gcsel}).~The ellipticity criteria are based on Local Group GCs \citep[see also][]{jordan09}, while the FWHM cut is set at about $\sim\!10\times$ the one of the stellar PSF, and in \cite{brescia12} we showed that using more restrictive criteria may result in losing extended GCs, such as $\omega$Cen. The photometric uncertainty cut is to ensure reliable fitting \citep[approximately equivalent to][]{paolillo11}, since at less conservative cuts the galaxy background begins to dominate (see Section~\ref{ln:bckg} and Figure~\ref{gclf}).~Additional criteria are the {\sc SExtractor} flag parameter set to $<\!4$, which excludes objects with incomplete and/or corrupted photometry apertures that are very close to the frame edges, and the total isophotal area limit of $\la\!6000$ pixel\footnote{We note that in \cite{paolillo11} and \cite{brescia12} the selection criteria were somewhat different, although broadly consistent, as those works had a different objective.}, which eliminates particularly extended galaxies and saturated foreground stars.

The final input catalog contains 6634 sources.~We show the $V_{\rm F606W}$ luminosity function of all detected and selected objects in Figure~\ref{gclf}.~We point out that the above selection criteria serve only as preparation of our sample for the next step of the analysis, i.e.~the profile fitting routine and are intended to minimize human interaction during the fitting process. In particular, they do not affect our final results.

\subsection{Estimating the Background Galaxy Contribution}
\label{ln:hudf}
We estimate the contribution of the background galaxy population to the luminosity distribution of our input catalog (see Fig.~\ref{gclf}) by applying the exact same photometry procedure to the F606W observations that were obtained as part of the Hubble Ultra-Deep Field (HUDF) program \citep{beckwith06}.~The HUDF observations were conducted in 112 sub-exposures spread over 56 orbits with a total integration time of 135320 seconds and give us excellent access to high-quality background galaxy photometry.~We obtained the drizzled HUDF image and the corresponding weight map from the Hubble Data Archive as reduced higher-level science products\footnote{http://archive.stsci.edu/prepds/udf/udf\_hlsp.html}, which were produced with virtually identical multidrizzle parameters compared with our procedure (see Sect.~\ref{ln:drizz}). To avoid unnecessary profile fitting of the many extended and amorphous sources in the HUDF we use SExtractor to measure their MAG\_BEST magnitudes, corrected for Galactic foreground extinction $E_{(B-V)}\!=\!0.008$ mag, and plot the corresponding luminosity function in Figure~\ref{gclf}.~Since the final HUDF frame covers 11 arcmin$^2$ with a spatial sampling of 0.03\arcsec/pix, we, therefore, scale the galaxy background number counts by a factor of 9.005 to match the survey area of our ACS mosaic.

The plot shows the remarkable similarity of the faint-end of the luminosity distribution of our sample with the background galaxy population, modulo a small difference at faint magnitudes $V_{\rm F606W}\!\ga\!26$\,mag, which is likely the manifestation of cosmic variance; e.g.~there is a variation in the number of background galaxy clusters in our ACS mosaic field.~This is consistent with the results of \cite{hilker99} and \cite{drinkwater00} who find a background galaxy cluster at $z\!=\!0.11$ behind the core of the Fornax galaxy cluster.

\section{Analysis}

\label{ln:sp}
At the distance of Fornax ($20.13\pm0.4$ Mpc) one arcsecond spans 97.6 pc.~On our drizzled ACS frames one pixel with the angular size of 0.03\arcsec\ subtends therefore 2.93 pc at the distance of NGC\,1399.~This is similar to the typical half-light radius for Milky Way globular clusters \citep{harris96}.~HST's confusion limit, $\delta$, at the pivot wavelength of the F606W filter, $\lambda_p\! =\! 606$ nm, can be estimated via $\delta\! =\! 1.22\, \lambda_p/D$, where $D\!=\!2.4$ meter of the HST primary mirror.~We obtain $\delta\!=\!6.2$ pc at the distance of Fornax.~However, because we are fitting analytical, multiple-component 2-D surface brightness profiles, our nominal spatial resolution is much better than the computed confusion limit.~The exact numerical value of the spatial resolution limit is determined through detailed artificial cluster experiments, which are discussed in Section~\ref{ln:art} in detail.

Observations of the integrated-light profile $\Sigma(\rvec)$ of resolved astronomical objects measure their surface brightness variations $\mu(\rvec)$ over the 2-D spatial extent ($\vec{r}\!=\!\rvec$ for spherically symmetric sources) convolved with the instrumental point-spread function $P(\rvec)$ and the detector diffusion kernel $D(\rvec)$, plus, in the simplest case, an additive noise term $N(\rvec)$:
\begin{equation}
\label{eq:obs}
\Sigma(\rvec)=2\pi\int\limits_{r_1}^{r_2}
\left\{\mu(\rvec) \otimes P(\rvec) \otimes D(\rvec) 
+\! N(\rvec)\right\}\rvec d\rvec
\end{equation}
where $\mu(\rvec)$ is the sum of the source and background surface brightness $\mu_{\rm s}(\rvec)\!+\! \mu_{\rm b}(\rvec)$.~The access to surface brightness profiles of distant objects (e.g.~globular clusters in NGC\,1399) is therefore limited by the spatial resolution of the data (i.e.~the width of functions $P(\rvec)$ and $D(\rvec)$), the brightness of the sky (i.e.~where $\mu(\rvec)\!\approx\!\mu_{\rm b}$), and the noise properties of the data (i.e.~$N(\rvec)$).~Among today's imaging instruments that operate at optical wavelengths, the ideal case of $P(\rvec)\otimes D(\rvec)\!\rightarrow \!\delta(\rvec)$ and $N(\rvec)\!\rightarrow\!0$ is best approximated by HST.~In particular, the ACS/WFC camera provides a large field of view ($\sim\!202\arcsec\!\times\!202\arcsec$) over which the geometric variations of $P(\rvec)\otimes D(\rvec)$ are relatively stable and well understood \citep{anderson05, jee07}.~An additional major advantage of HST observations is the very low sky background with a typical surface brightness $\mu_{{\rm b}, V}\!\ga\!22.5$ mag arcsec$^{-2}$ (see also ACS Instrument Handbook).

\subsection{King Surface Brightness Profile}
The reason for the great success of the King profile in parametrizing the surface brightness profiles of most Galactic globular clusters is their structural homology, and is a simple consequence of the fact that virtually all of these systems have ages far in excess of their relaxation times \citep[e.g.][]{king68, ill76, dacosta79, kukarkin79, chun80, trager95}.~We note here {\it en passant} that this might not be the case for more extended sources \citep[e.g.][]{misgeld11}.~The King profile \citep{king62}, which is defined as
\begin{equation}
\label{eq:king}
\mu_{\rm K}(\rvec)=k\left[\left(1+\frac{\rvec^2}{r_c^2}\right)^{-\frac{1}{2}}\! - 
                           \left(1+\frac{r^2_t}{r_c^2}\right)^{-\frac{1}{2}}\right]^2,
\end{equation}
describes the surface number density in the range $0\leq |\rvec| \!<\! r_t$ and is zero for $\rvec\! \geq\! r_t$.~Its shape is governed by the core radius $r_c$, at which the projected surface density is half the central stellar surface density, which itself is set by the cluster gravitational binding energy ($r_c\!\approx\!3\sigma /\sqrt{4\pi G\rho_o}$ for $r_t/r_c\!\gg\!1$, see e.g.~\citealt{binney87}).~If the GC is tidally filling, $r_t$ can be considered the tidal radius, otherwise $r_t$ marks the limiting radius beyond which the stellar density drops to zero; this is sometimes referred to as the King radius ($r_k$).~The profile is normalized to the central surface brightness by $k\!=\!\mu(0)(1\!-\!1/\sqrt{1\!+\!r^2_t/r_c^2})^{-2}$.~The family of King profiles is parametrized by the concentration $c\!=\!r_t/r_c$, which is directly proportional to the central potential $W_o$ via $c\simeq 9.12\! +\! (W_o\!-\!4.215)^{3.064}$ for $W_o\leq12$ (\citealt{king66}, see also \citealt{binney87}). The basic assumption of this parametrization is a truncated (so-called "lowered") Maxwellian phase-space distribution of GC member stars in addition to the premise of orbital isotropy.

It is assumed that the King profile is a valid description of the surface-brightness profiles of extragalactic GCs \citep[e.g.][]{harris02, harris10, sharina05, jordan05, huxor05, gomez06, barmby07, mclaughlin08, masters10}. In other words, we assume a universal homology among globular clusters and adopt the King structural parameters as a sufficient set to describe their light profiles.~However, we have to keep in mind that such objects may not be well represented by isotropic, single-mass, isothermal spheres but may be better described by other profiles. \cite{mclaughlin05} show that other profiles such as the \cite{wilson75} profile or power-law profiles \`{a} la \cite{elson87} fit the outermost parts of Milky Way and Magellanic Clouds GCs as well or better than classic King profiles.~Furthermore, \cite{webb12a} compare King62 models fits to King66 \citep{king66}, Wilson75 \citep{wilson75}, and S\'ersic models \citep{sersic68} for GCs in M87, and find that King66 models significantly underestimate cluster sizes, while Wilson75 fits are in close agreement with King62 measurements.~However, we keep in mind that GCs outside the Local Group may have experienced different dynamical evolution histories given that their host galaxies may have undergone more violent merging and accretion histories \citep[e.g.][]{baumgardt03} that may give rise to a larger variety of unusual GC surface-brightness profiles.~Our analysis will necessarily be less sensitive to the outer low-surface brightness outskirts of the NGC\,1399 GCs than to their half-light or core properties.~Since all the aforementioned profiles are virtually identical in their inner parts \citep[i.e.~within their half-light radius, see][]{mclaughlin05} we adopt the King62 profile for the rest of the analysis.~The main reason is that for marginally resolved GCs the profile choices become rather unconstrained and more complex models often diverge or give degenerate results \citep{barmby07, mclaughlin08, harris10}, whereas the King62 profile provides the most robust measures of GC structural parameters for both marginally resolved and well resolved targets.

\subsection{The Fitting Routine}
\label{ln:profilefit}
To derive the structural parameters of NGC\,1399 GCs we fit their surface brightness profiles using a modified version of the GALFIT package that includes the King profile as a fitting option \citep[v3.0,][and references therein]{cpeng10}.~Previous software packages such as {\sc ishape} \citep{larsen99}, {\sc gridfit} \citep{mclaughlin05} and {\sc kingphot} \citep{jordan05} offer valid alternatives for measuring GC structural parameters.~However, {\sc ishape} generally uses fixed King concentration parameters and deals with elliptical sources in a semi-analytical way.~These three routines do not allow for flexible fitting of multiple blended sources with various profile types plus a variable background component.~Additional advantages of our code is the execution handling and speed, which allows us to efficiently conduct large amounts of artificial cluster experiments (see below).

\begin{figure*}[!t]
\centering
\includegraphics[bb=0 -3 30 30, width=5.2cm]{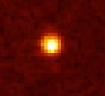}
\includegraphics[bb=0 -3 30 30, width=5.2cm]{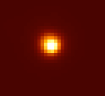}
\includegraphics[bb=0 -3 30 30, width=5.2cm]{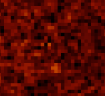}
\includegraphics[width=5.2cm]{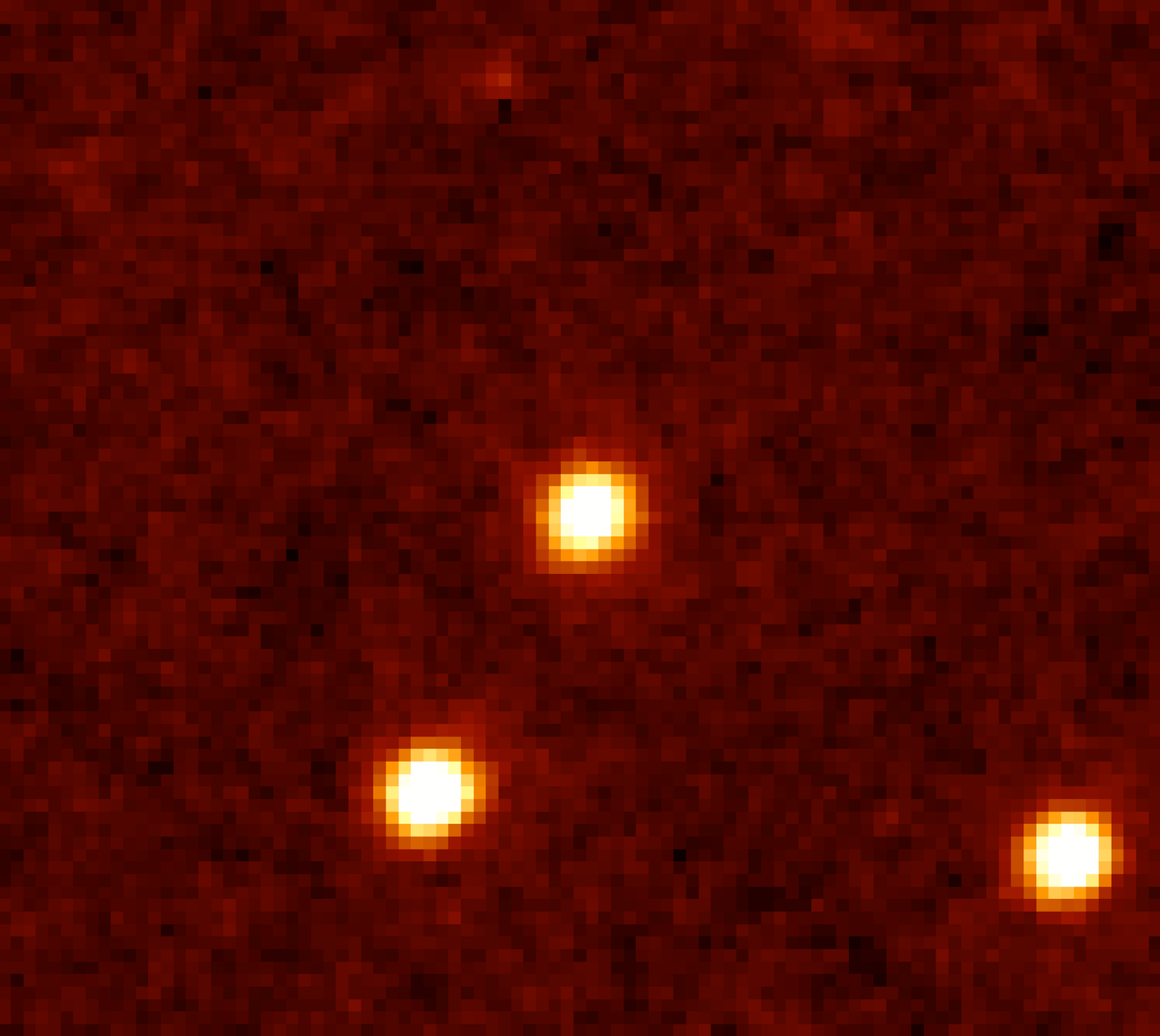}
\includegraphics[width=5.2cm]{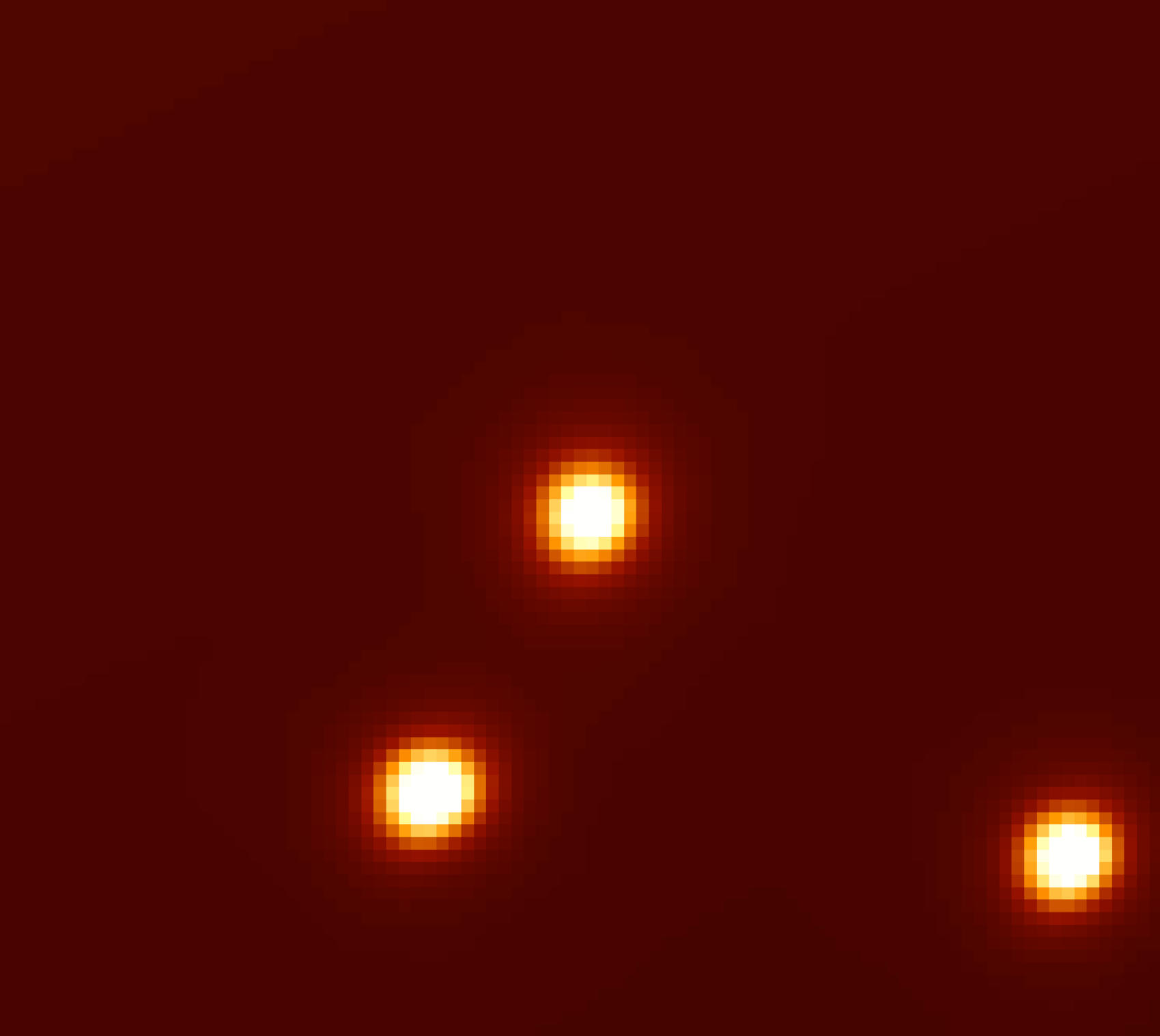}
\includegraphics[width=5.2cm]{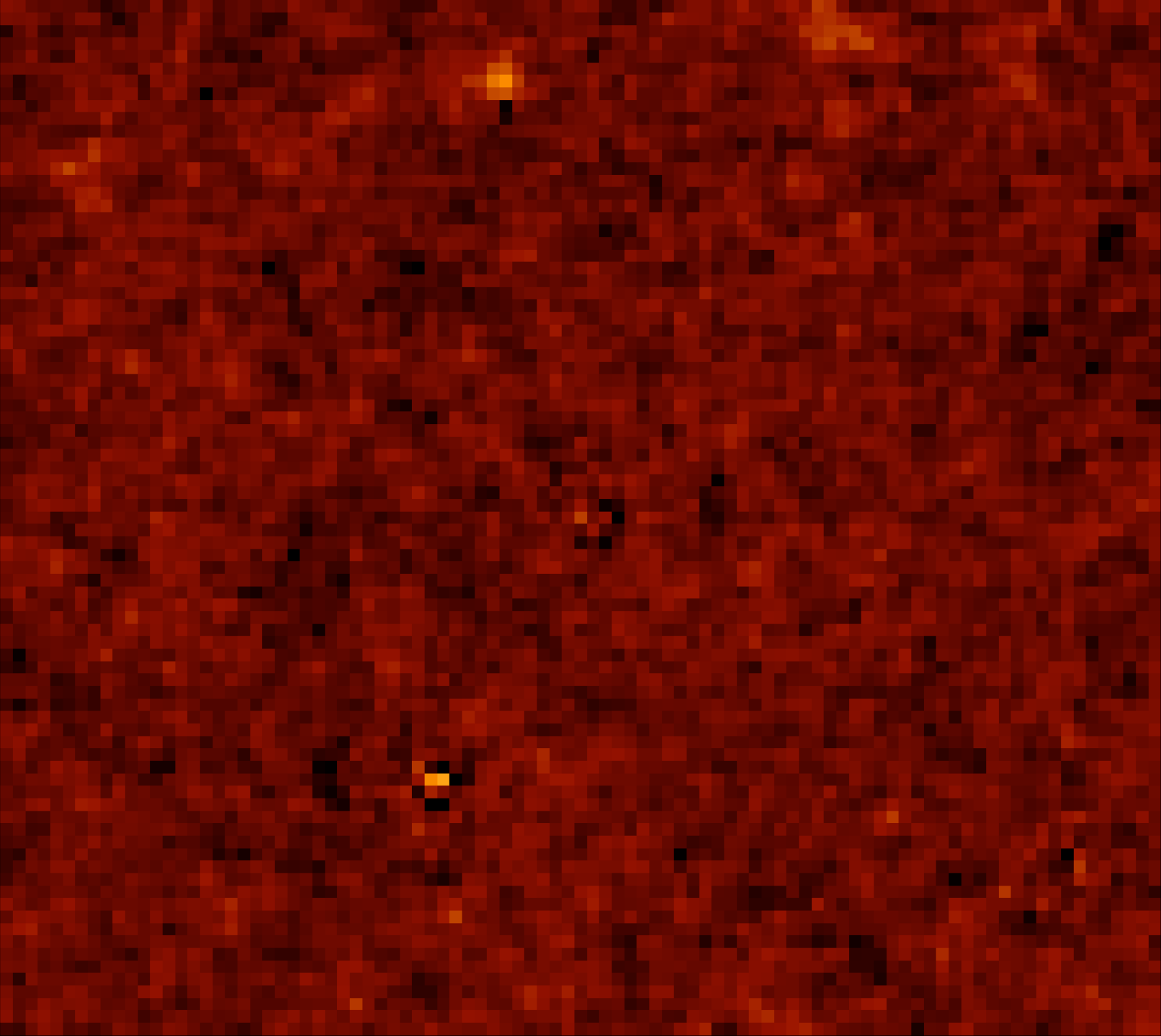}
\caption{Postage stamp cutouts of typical examples of King profile fits from our fitting routine. The left panels show the data and the centre panels illustrate the models, both shown with the same color stretch. The right panels are the residual maps with $10\times$ larger color stretch to accentuate the residual noise, which is less than
$\sim1\%$ for these particular cases. The field of view in the top-row panels is $\sim1\arcsec\!\times\!1\arcsec$, while the imaged area in the bottom-row panels is $\sim2.5\arcsec\!\times\!2.5\arcsec$.}
\label{pdf:fit}
\end{figure*}

We account for blended sources within the fit region of each GC and simultaneously fit profiles to sources which are less than five magnitudes fainter than the target within a radius of two FWHM, and to sources less than three magnitudes fainter outside this region.~At the same time, we match the contributions of the sky+galaxy surface brightness by fitting a surface within the same area.~The code uses a $\chi^2$ minimization scheme to simultaneously optimize the fit to each source and the local background surface brightness. Because some objects are blended with nearby very extended sources, we additionally use various profiles types for those blended objects, such as clearly extended nearby dwarf galaxies for which we choose the S\'{e}rsic profile \citep{sersic68}. Extended objects that have isophotal areas $\ga\!10\times$ larger than the fitting area are well approximated by a simple sloped background contribution.~A representative example of the fit quality for a typical GC in NGC\,1399 is shown in Figure~\ref{pdf:fit}.

\subsection{Constructing the PSF Library}
Equation~\ref{eq:obs} shows that detailed knowledge of the local PSF over the entire image is mandatory to obtain meaningful measurements of profile parameters, and the most realistic representation of the convolution product $P(\rvec)\otimes D(\rvec)$ is provided in form of a library of empirically measured PSFs \citep[see discussion in][]{georgiev09a}.~Such a library of {\it effective} PSF (ePSF) profiles based on repeated ACS observations of dense stellar fields was presented for several HST/ACS filters by \cite{anderson05} and \cite{anderson06}. Because of the fully empirical approach to build such a library \citep{anderson00}, this collection provides the best characterization of the ACS/WFC-PSF for our purposes, as it preserves the variations of high {\it and} low-contrast features of the PSF with high spatial on-chip sampling.~This is superior to the PSF modeling techniques provided by the TinyTim simulator\footnote{http://www.stsci.edu/software/tinytim/tinytim.html} and other parametric PSF approximations \citep{jee07}, as well as building the PSF library from the science images themselves where the relative foreground stellar density is not sufficiently high to obtain a clean PSF star sample.

The ePSF library provides a set of $10\!\times\! 9$ PSF profiles uniformly covering the WFC field of view.~Each ePSF is oversampled by a factor of four to account for shifts of the source centroid with respect to the pixel center and applies only to the individual distorted ACS exposures ("flt" files).~In order to transform the ePSFs into the final drizzled images, we need to apply our data reduction process to the library itself.~To this end we designed a custom software package \citep[{\sc MultiKing\footnote{The IDL source code to produce the drPSF library grid images is available at http://people.na.infn.it/$\sim\!$paolillo/Software.html.}}, see][]{paolillo11} to overlay the Anderson PSF grid onto a set of empty WFC frames, reproducing the actual data frame properties (orientation, dither pattern, astrometry, etc.).~The grid positioning was modified on each frame to preserve the sky coordinates of each PSF, properly accounting for geometric distortions that affect the WFC "flt" frames, as would be expected for a real source within a set of observations taken with our dithering pattern. Since each dither pattern is executed with slightly varying sub-integration pointings, this procedure was applied to each individual pointing of the ACS mosaic. Finally, the dithered ePSF frames were combined together in the same way as the science frames, producing a {\it drizzled effective PSF} (drPSF) library for each individual ACS tile. The specific stellar PSF at a random location within our final images is chosen to be the nearest drPSF within the template grid.~We use these drPSF libraries for the subsequent analysis. Our code was already implemented in the study of \cite{goudfrooij12} who successfully used the drPSF approach to measure star cluster sizes in NGC\,1316.

\subsection{Artificial Cluster Experiments}
\label{ln:art}
\begin{figure*}[!ht]
\centering
\includegraphics[width=7.4cm]{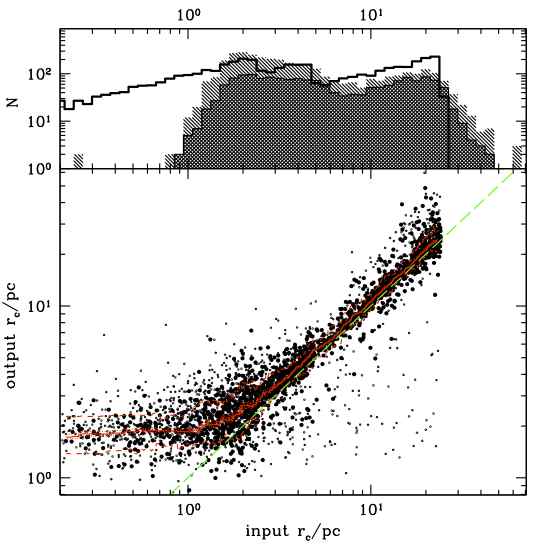}
\includegraphics[width=7.4cm]{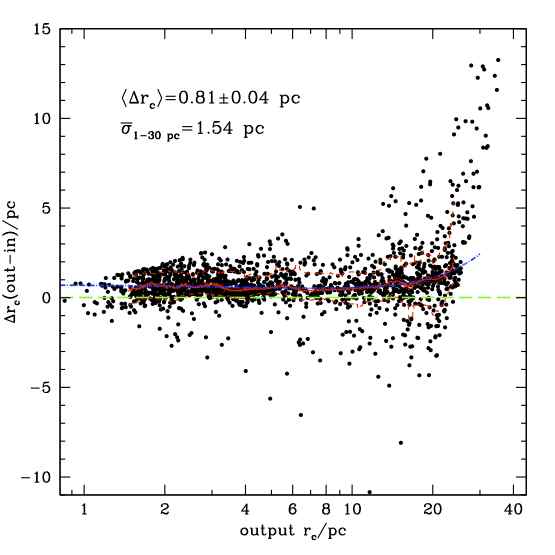}
\includegraphics[width=7.4cm]{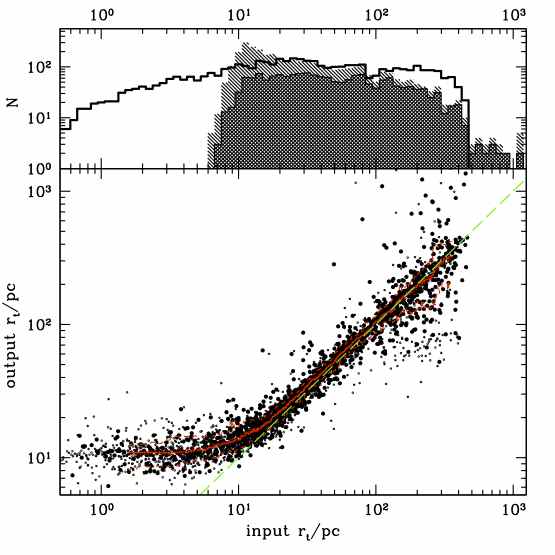}
\includegraphics[width=7.4cm]{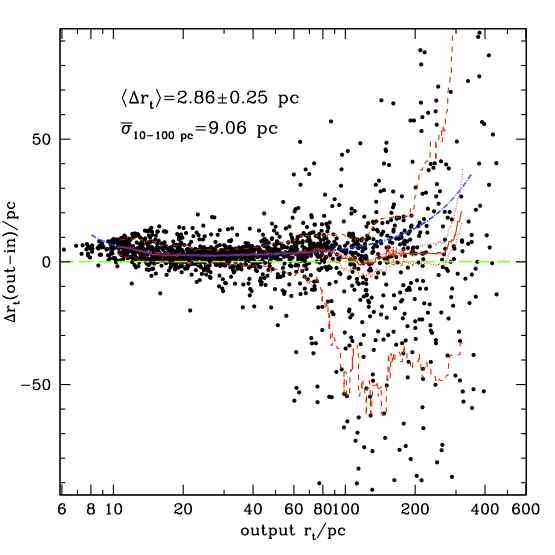}
\includegraphics[width=7.4cm]{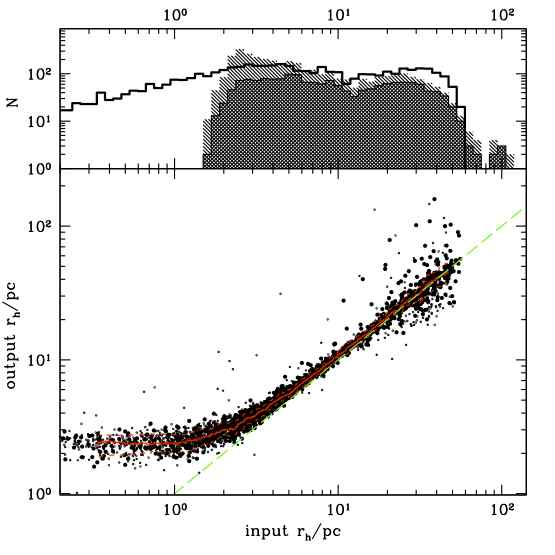}
\includegraphics[width=7.4cm]{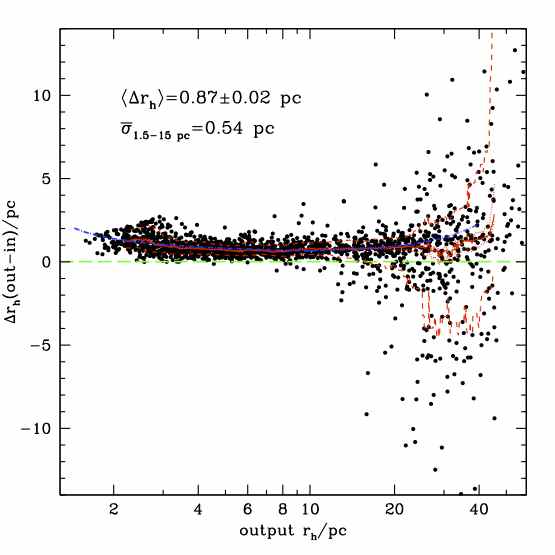}
\caption{The figure shows the recovery quality of King profile structural parameters ($r_c$, $r_t$, $r_h$) performed with artificial star cluster. The left panel column shows the input vs.~output parameter values and is divided in low-quality ({\it small dots}: ${\hat\chi}^2\!>\!1$) and high-quality profile fits ({\it large dots}: ${\hat\chi}^2\!\leq\!1$). A dashed line indicates the equality relation and a solid red curve is a sliding-median probability density estimate with its 1-$\sigma$ limits ({\it dashed curves}) and the error of the mean ({\it dotted curves}). The histogram sub-panels show the input parameter distribution ({\it open histogram}) in comparison with the high-quality ({\it double-shaded histogram}) and low-quality profile fits ({\it single-shaded histogram}). The right column of panels shows the corresponding residual functions for each structural parameter as a function of the output value. Blue dash-dotted curves are correction functions that are fits to the data (see text for details).}
\label{gc_sim}
\end{figure*}

Every attempt to determine the structural parameters of extragalactic GC is affected by measurement uncertainties, parameter covariance, and other inherent systematic characteristics of the dataset and measuring technique.~To test the robustness of our measurements (under the assumption that the King62 profile describes the NGC\,1399 GC profiles sufficiently well) and probe parameter correlations and systematics we used our {\sc MultiKing} code to create and add artificial star clusters to our ACS science frames and attempt to recover their structural parameters with our profile fitting routines using the exact same approach as for the analysis of NGC\,1399 GCs.~This process includes convolving the appropriate drPSFs of the corresponding ACS tile with King profiles of varying structural parameters and inserting the noise-corrected clusters at random locations in the eastern, southern, and central tile of the ACS mosaic. In this way we include 1500 artificial clusters per tile in 15 runs each to avoid effects of artificial crowding. The input structural parameters cover a broad dynamic range that aims to sample crucial values around the resolution and confusion limits more densely. In particular, it covers the typical sizes of Galactic and LMC globular clusters.

The recovery quality of the core radius, $r_c$, half-light radius, $r_h$, and tidal radius, $r_t$, is illustrated in Figure~\ref{gc_sim}. For each parameter we show the input vs.~output correlation, together with a ''sliding-median'' probability density estimate and the corresponding 1-$\sigma$ contours as well as the error-of-the-mean margin. The renormalized profile fit quality serves as a metric to divide our artificial cluster sample in low- and high-quality fits, the division of which is done at the renormalized reduced chi-square ${\hat\chi}^2\!=\!1$.~This division generally corresponds to faint and bright sources. The corresponding histograms in the left panels of Figure~\ref{gc_sim} compare the input with the recovered parameter distribution and indicate biases in our measuring process.~Our cluster experiments are consistent with the results presented in \cite{carlson01}.~In particular, all our bona-fide sample GCs have an integrated S/N~$\ga\!100$ in agreement with the minimum prescription of \citeauthor{carlson01} to measure sizes of marginally resolved GCs\footnote{We also note that \cite{carlson01} claim that S/N~$>500$ is required in order to fully recover all King model parameters for every type of GC out to a distance of $\sim\!40$\,Mpc, i.e.~twice as far as NGC\,1399. On the other hand, they state that S/N~$\!\approx\!100$ is appropriate for, e.g.~Virgo galaxies, or less concentrated systems, and that GC half-light radii are recovered with even better accuracy.~Furthermore our spatial sampling (pixel size) is $\sim\!3$ times better than what was used in their study.}.~In the following, we discuss and quantify these systematics to provide numerical estimates of the reliability of the subsequent structural parameter analysis.

\subsubsection{The Recovery Fidelity of the King Core Radius}
\label{ln:fidel_c}
The top panels in Figure~\ref{gc_sim} show how our code recovers the King core radius, $r_c$.~From the left panel, i.e.~input vs.~output $r_c$ diagram, it is evident that the spatial resolution of our dataset becomes increasingly poorer at $r_c\la3$ pc, and we see nicely a ''leveling off'' of the relation towards smaller spatial scales. The reader should be aware that the logarithmic scaling in this plot is chosen to show exactly this physical limit and exaggerates this effect optically. In the right panel we plot the $r_c$ difference relation, in the sense $\Delta r_c = r_{c,{\rm out}} - r_{c,{\rm in}}$, vs.~the recovered core radius $r_{c,{\rm out}}$. The graph illustrates that the measurements can be robustly corrected with a small systematic offset of the form $\langle\Delta r_c\rangle = 0.81\pm0.04$ pc with a mean standard deviation of $\bar{\sigma}_{\rm 1-30 pc}=1.54$ pc, which is equivalent to the overall $r_c$ measurement uncertainty in this range. While the $\Delta r_c$ trend around the spatial resolution limit allows an almost linear correction it is clear that the scatter in $\Delta r_c$ increases towards larger $r_c$, which is due to the confusion limit of the data, e.g.~blended sources, sky background fluctuations, etc.~At this end we see a higher-order systematic trend that cannot be approximated with a simple offset.~We, therefore, use the correction function
\begin{equation}
\label{eqn:rc}
\phi_{r_c} = 0.731 - 5.563\cdot10^{-2} r_{c,{\rm out}} 
+ 3.742\cdot10^{-3} r_{c,{\rm out}}^2
\end{equation}
to fit the overall trend in $\Delta r_c$ as a function of $r_{c,{\rm out}}$ and correct our $r_c$ measurements for $r_{c,{\rm out}}\!\in\![1,30]$ pc. The function is shown as dash-dotted curve in the upper right panel of Figure~\ref{gc_sim} and approximates the probability density curve very well in the range $r_c\!\approx\!2\!-\!20$ pc, which we consider as our high-confidence range for the King core radius measurements.

\subsubsection{The Recovery Fidelity of the King Tidal Radius}
\label{ln:rhcorr}
The tidal radius, $r_t$, probes the outskirts of the GC light distribution.~Our tests recover $r_t$ with good accuracy in the range between $\sim\!10$ and 100 pc (see middle panels in Fig.~\ref{gc_sim}). The mean residual is $\langle\Delta r_t\rangle\!=\!2.86\pm0.25$ pc with an average standard deviation of $\bar{\sigma}_{\rm 10-100 pc}=9.06$ pc.~The lower limit is set by the starting value of our fitting routine which is ten times the initial core radius value, so that some clusters with a large core radius and a slightly larger tidal radius end up with an overestimated tidal radius, because the numerical convergence of the code for fits with very similar core and tidal radii is internally defined by the core radius. Note, that the tidal radius cannot be smaller than the core radius. For objects more extended than $\sim\!100$ pc we run into background confusion and fit degeneracy problems, introduced by nearby diffuse galaxy components and satellite objects (which are fit as described in Sect.~\ref{ln:profilefit}). Hence, the fits become poorly defined beyond such large tidal radii, simply because there is not enough signal-to-noise in the low surface-brightness wings of the profiles.~We approximate the corresponding residual trend with the following two component correction function
\begin{equation}
\label{eqn:rt}
\phi_{r_t}= 
\begin{cases} 
0.646 + 22.458\, (r_t\!-\!5.581)^{-0.86} 
	& \hspace{-0.15cm} \text{if $r_t\!\in\! [10,25)$,} \\
2.078 + 6.675\cdot10^{-3}\, (r_t\!-\!1.11) + \\
+ 2.591\cdot10^{-4}\,(r_t\!-\!1.11)^2 
	& \hspace{-0.15cm} \text{if $r_t\!\in\![25,100]$,}
\end{cases}
\end{equation}
which is valid in $r_t\!\in\![10,100]$ and robustly follows the probability density estimate out to the extreme edges of the parameter range.
	
\subsubsection{The Recovery Fidelity of the Half-Light Radius}
\label{ln:rhfidelity}
The GC half-light (or effective) radius, $r_h$, is a structural parameter that emerges from the correlation of the King core and tidal radius, as described by Equation~\ref{eq:king} and encircles 50\% of the total GC light.~The half-light radius is relatively stable throughout the GC dynamical evolution and is predicted to evolve much slower with time ($r_h\!\propto\! t^{2/3}$) than the tidal and core radius \citep{henon73, henon75, elson87rev, murphy90, murray92}.~One major advantage of $r_h$, is its relatively effortless accessibility in more distant stellar systems and because of its slow evolution it provides the most reliable measure of the true size distribution function of extragalactic GC system.~Formally, the GC half-light radius, $r_h$, is defined as
\begin{equation}
\label{rhdef}
2\pi\int\limits_{0}^{r_h}\mu(\rvec)\rvec d\rvec = 
\pi\int\limits_{0}^{\infty}\mu(\rvec)\rvec d\rvec\; ,
\end{equation}
and can be evaluated with the integral form of the King profile which can be written as
\begin{eqnarray}
\label{kingint}
\lefteqn{2\pi\int\limits_{r_1}^{r_2}\mu_{\rm K}(\rvec)\rvec d\rvec = \left[
\frac{kr_c}{\varrho}\left\{\varrho\,\mbox{arctan}\left(\frac{\rvec}{r_c}\right) 
+ \right. \right.} \\
&\left. \left. r_c\left(\rvec - 2\sqrt{\varrho}\, 
\ln\left(2\rvec\sqrt{\varrho} + 2r_c^2\sqrt{1 + \frac{r_t^2}{r_c^2}} 
\sqrt{1 + \frac{\rvec^2}{r_c^2}}\right)\right)\right\}\right]_{r_1}^{r_2} \nonumber
\end{eqnarray}
where $\varrho=r_c^2 + r_t^2$. Since the half-light radius $r_h$ cannot be written in a closed analytic form from Equations~\ref{rhdef} and \ref{kingint}, it has to be evaluated numerically. Hence, we determine $r_h$ from the direct numeric integration of the King profile for each individual cluster and, thus, probe immediately the influence of parameter correlations between $r_c$ and $r_t$ on the integrated luminosity. The bottom panels of Figure~\ref{gc_sim} show that the average bias of the half-light radius is $\langle\Delta r_h\rangle\!=\!0.87\pm0.02$ pc with an average standard deviation of $\bar{\sigma}_{\rm 1.5-15 pc}\!=\!0.54$ pc.~This is in excellent agreement with the results of \cite{harris09} who found $\sigma_{r_h}\!=\!1.1$ pc as mean uncertainty for size measurements of GCs at the distance of $\sim\!40$ Mpc, based on similar data of distant BCGs which are roughly twice as far away as NGC\,1399.~The reduced mean uncertainty of $r_h$ is likely due to parameter correlations between $r_c$ and $r_t$, the uncertainties of which compensate each other to leave $r_h$ a very reliable parameter of GC size. The results of our artificial cluster experiments indicate that we can measure {\it and} correct $r_h$ reliably for $r_h\in[1.5,19]$ pc.~We compute the corresponding correction function of the form
\begin{equation}
\label{eqn:rh}
\phi_{r_h}= 
\begin{cases} 
0.33 + (r_h\!-\!0.354)^{-4.26} 		& \text{if $r_h\!\in\![1.5,7)$,} \\
0.449 - 5.766\cdot10^{-2}\, (r_h\!-\!0.1) + \\
+ 5.869\cdot10^{-3}\, (r_h\!-\!0.1)^2	& \text{if $r_h\!\in\![7,19]$.}
\end{cases}
\end{equation}

In summary, we now understand the fidelity and limitations of our structural parameter measurements and move on to test the influence of the variable background surface brightness in our ACS mosaic.

\subsubsection{The Influence of the Variable Galaxy Background}
\label{ln:bckg}
\begin{figure}[!t]
\centering
\includegraphics[width=8.9cm]{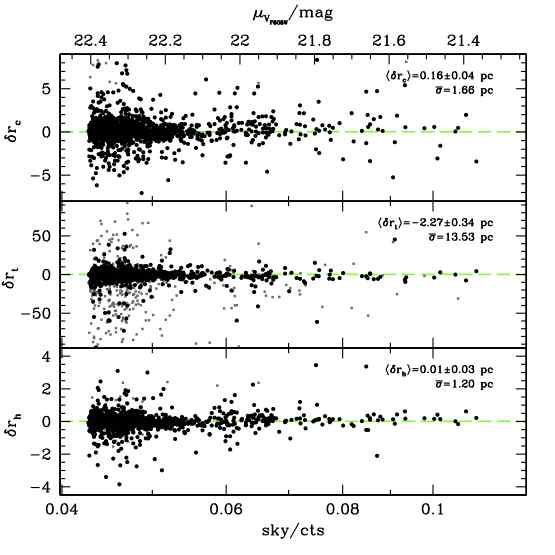}
\caption{Illustration of the residuals around the correction functions (see Equations~\ref{eqn:rc}, \ref{eqn:rt} and \ref{eqn:rh}) as a function of background surface brightness.~High- and low-quality fits are depicted as grey and black dots, respectively, and are defined as fits with a reduced $\chi^2$ below and above unity. Mean residuals and dispersion are given in each panel. \cite{caon94} measures the $B$-band surface brightness profile of NGC\,1399 out to $\sim\!15\arcmin$ galactocentric radius. We use the numbers from \cite{sandage75} who reports $B\!-\!V\!\simeq\!0.95$ to $0.98$ mag in the range $54.3\arcsec\!- 107.8\arcsec$ from the center of NGC\,1399 to obtain a rough estimate of the $V$-band surface brightness.}
\label{pdf:residbkg}
\end{figure}
After correcting for biases in our measuring procedure we explore in the following the influence of the variable galaxy surface brightness in the studied field on our structural parameter measurements.~To do so we compute the residuals with respect to the correcting functions $\phi_{r_i}$ in Figure~\ref{gc_sim} in the form $\delta r_i\!=\! (r_{i,{\rm out}}\! -\! r_{i,{\rm in}}) - \phi_{r_i}$ where the index $i$ stands for the core, tidal, and half-light radius, respectively. The residuals are then plotted in Figure~\ref{pdf:residbkg} as a function of the background counts. All our measurements are shown in this figure, however, only the values within the confidence limits of Equations~\ref{eqn:rc}, \ref{eqn:rt} and \ref{eqn:rh} are considered in the computation of the mean residual statistics.~This exercise shows that our GC profile fitting routine accounts very robustly and without any significant residual systematic for the variable background.~Our tests probe surface brightness levels fainter than $\mu_{V_{\rm F606W}}\ga21.4$ mag, which corresponds to galactocentric radii $r\ga 30\arcsec$, and we expect that the final measurements are reliable without further corrections to within the quoted uncertainties of our artificial cluster experiments within the above $\mu_{V_{\rm F606W}}$ range.

\subsection{Comparison with ACSFCS Measurements}
\label{ln:litcomp}
\begin{figure}[!b]
\centering
\includegraphics[width=8.9cm]{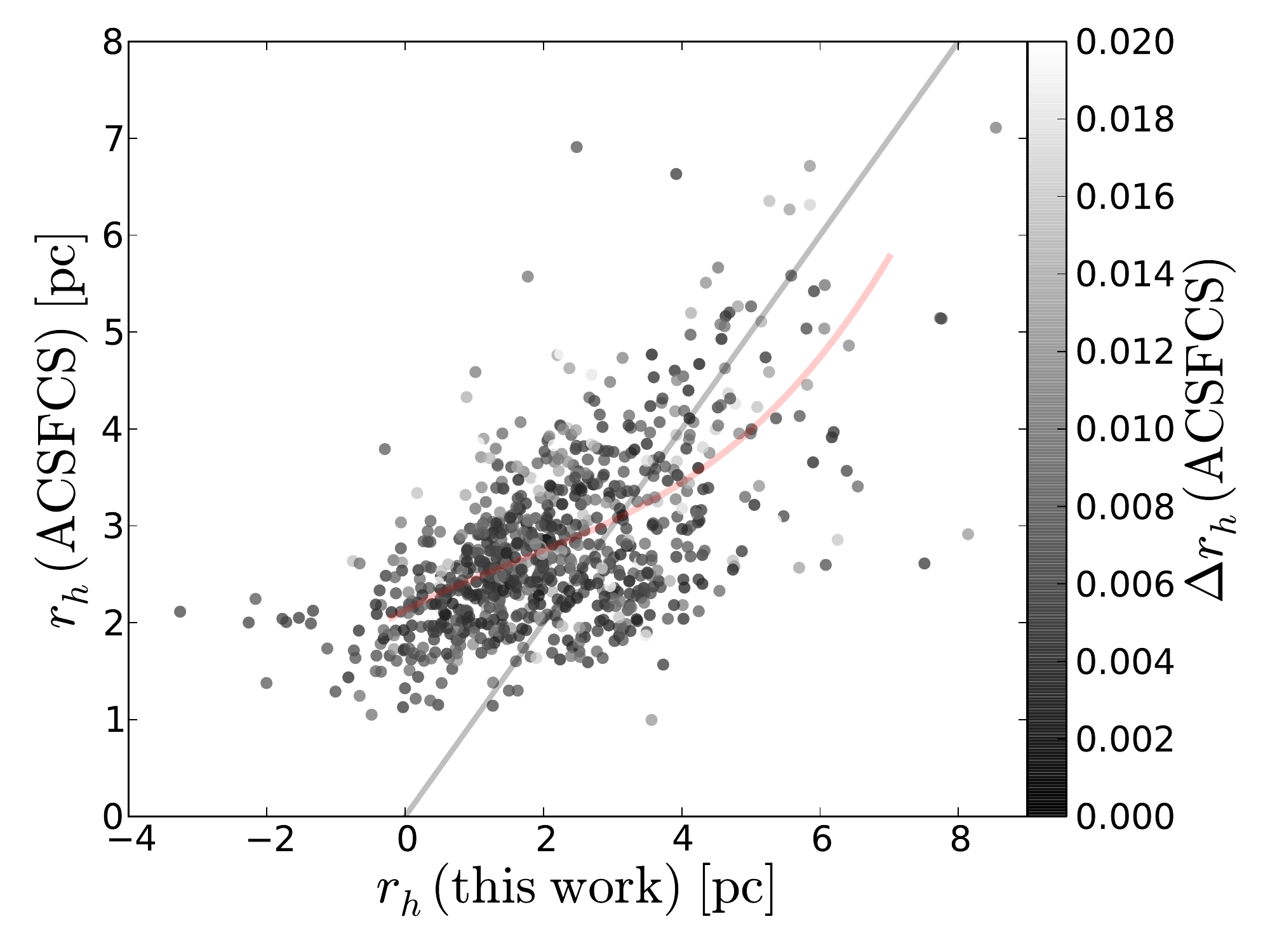}
\caption{Comparison of GC half-light radius measurements from this work and the ACS Fornax Cluster Survey \citep[ACSFCS, see][]{jordan07, masters10} for the same target GCs in NGC\,1399. The grey line shows the one-to-one relation. The greyscale parametrizes the ACSFCS measurement uncertainties, $\Delta r_h$, which are computed as the square root of the square sum of the individual uncertainties in the F475W and F850LP filters.~The red line is a third-order polynomial approximation to the data.}
\label{pdf:litcomp}
\end{figure}
In the following we compare our measurements to the most recent GC half-light radius measurements in NGC\,1399 based on the ACS Fornax Cluster Survey \citep[ACSFCS, see][]{jordan07} which observed the galaxy with one central pointing.~The ACSFCS GC half-light radius measurements were conducted with the {\sc kingphot} software \citep{jordan05} and are restricted to the brightest GCs with $z\!\leq\!23.35$\,mag and colors $0.6\!\leq\!(g\!-\!z)\!\leq\!1.7$ mag. The ACSFCS data are comprised of $2\times565+90$\,sec exposures in F850LP and $2\!\times\!380$\,sec exposures in F475W. Compared to our $4\!\times\!527$\,sec, optimally dithered F606W observations, the ACSFCS data have, therefore, a somewhat lower S/N at an equivalent GC luminosity (due to lower system throughput in F475W and F850LP) and have a more sparsely sampled PSF due to their 2-step dither pattern \citep{jordan05, jordan07}.~We take the ACSFCS GC half-light radii published as part of the study presented in \cite{masters10} and use the arithmetic mean of their GC half-light radius measurements in the F475W and F850LP filters and select only GC candidates that were assigned a GC probability of $p_{\rm GC}\!\geq\!0.5$ \citep[see also][]{jordan09}.

Figure~\ref{pdf:litcomp} shows the direct comparison between the two samples where we find no significant offset beyond $r_h\!\approx\!2$ pc.~However, at smaller half-light radii, the influence of the correction function from Equation~\ref{eqn:rh} (see also Fig.~\ref{gc_sim}) becomes increasingly apparent as the ACSFCS data tend to be biased towards larger values relative to our measurements.~This is primarily due to the fact that the ACSFCS measurements are not corrected for measurement systematics by means of artificial cluster experiments as in our procedure (see Section~\ref{ln:rhfidelity}).~We parametrize the grey shading of data points in Figure~\ref{pdf:litcomp} with the measurement uncertainties, $\sigma_{r_h}$, of the ACSFCS data.~The main trend of the comparison is approximated by a third-order polynomial and depicts the shape of the correction function. The rms around this relation is 0.77 pc, and with our measurement uncertainty of 0.54 pc from the artificial cluster experiments described in section~\ref{ln:rhcorr}, we obtain $\Delta_{\rm total}\!=\!\sqrt{\sigma_{r_h}({\rm ACS})^2 - \sigma_{r_h}({\rm this\ work})^2}\approx 0.55$\,pc, which we regard as the total statistical uncertainty when comparing individual GC half-light radius measurements from various studies using different techniques.

\section{Results}

We use our surface brightness profile (SBP) fitting routine described in Section~\ref{ln:profilefit} to measure the structural parameters of all sources from our photometric input catalog (see Section~\ref{ln:photin}) and calibrate them with the correction functions $\phi_{r_i}$ derived in Section~\ref{ln:art} (see Equations~\ref{eqn:rc}, \ref{eqn:rt} and \ref{eqn:rh}).~Because of the higher measurement fidelity of the half-light radius, we use $r_h$ in the subsequent analysis and refer to it as GC size, unless stated otherwise.~We point out that for addressing other specific scientific topics, such as measuring the binary star formation efficiency via LMXB population analysis, other parameters such as the core radius and central surface brightness proved to be more diagnostic than $r_h$ \citep{paolillo11}.

\subsection{Total Object Magnitudes}

\begin{figure}[!t]
\centering
\includegraphics[width=8.9cm]{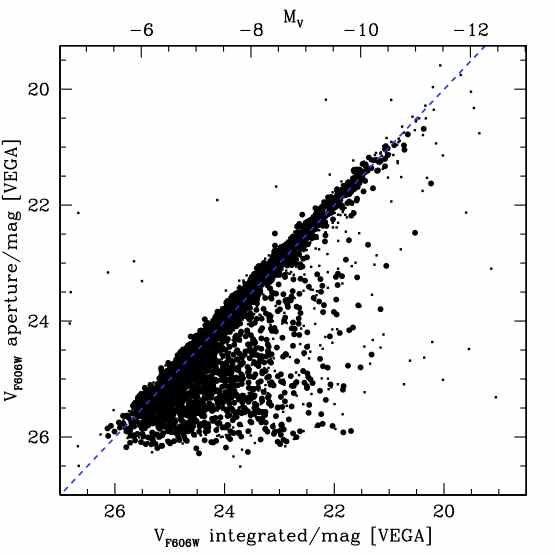}
\caption{Comparison of total object luminosities determined via aperture photometry and direct integration of their surface brightness profile. Both magnitudes are in the Vega system and are corrected for Galactic foreground reddening with $A_{\rm F606W}\!=\!0.038$ mag. The top ordinate indicated the absolute $M_V$ at the distance of Fornax. A blue dashed line shows the equality relation.}
\label{pdf:apint}
\end{figure}

To investigate correlations of structural parameters with GC brightness it is important to compute accurate total luminosities for our object sample.~In Section~\ref{ln:gcc} we corrected our aperture photometry with a generic aperture correction term to compensate for the light outside the $r\!=\!0.24\arcsec$ photometry radius which delivered the highest photometric S/N and served as a first guess for the structural parameter fitting routines. With our structural parameter measurements in hand we can now directly integrate the surface brightness profile of all targets and determine their total luminosities and compare them with the traditionally determined aperture magnitudes.~Figure~\ref{pdf:apint} shows the direct comparison of total Vega magnitudes measured via corrected aperture photometry and integrated SBP luminosities.~Small and large dots are defined as in Figure~\ref{gc_sim} for low and high-quality fits of the surface brightness profile.~Figure~\ref{pdf:apint} shows that the vast majority of our sample aligns very well with the one-to-one relation, which is due to the fact that most of our sample objects are marginally resolved GCs.~Clearly resolved objects scatter to the right of the one-to-one relation and have brighter integrated magnitudes and have too faint aperture photometry counterparts.~Their total aperture magnitudes at the GCLF turnover $M_V({\rm GCLF})\!\simeq\!-7.5$ mag are up to $\sim\!0.5\!-\!2$ mag fainter than the corresponding integrated SBP luminosities. In general, this is due to an average correction that is applied to all GCs when measuring GC luminosities via aperture photometry. This is direct evidence that for partially resolved and clearly resolved objects an average aperture correction term is not sufficient to determine their total luminosities. Note also that there are virtually no outliers left of the one-to-one relation which is visual assurance of our SBP fitting quality.

The study of \cite{kundu08} has previously claimed that certain GC parameter correlations, such as the color-luminosity relation for bright blue globular clusters may be the result of inappropriately applying average aperture corrections to multi-passband photometry object samples with widely varying structural parameters.~Although our structural analysis is based on the F606W filter only, to avoid such problems in what follows we use the directly integrated SBP magnitudes for the subsequent analysis, and point to the works of \cite{epeng09} and \cite{harris09} for a more detailed discussion of this filter-dependent aperture correction issue.

\subsection{Radial Velocity Information}
\label{ln:vrad}

\begin{figure}[!t]
\centering
\includegraphics[width=8.9cm]{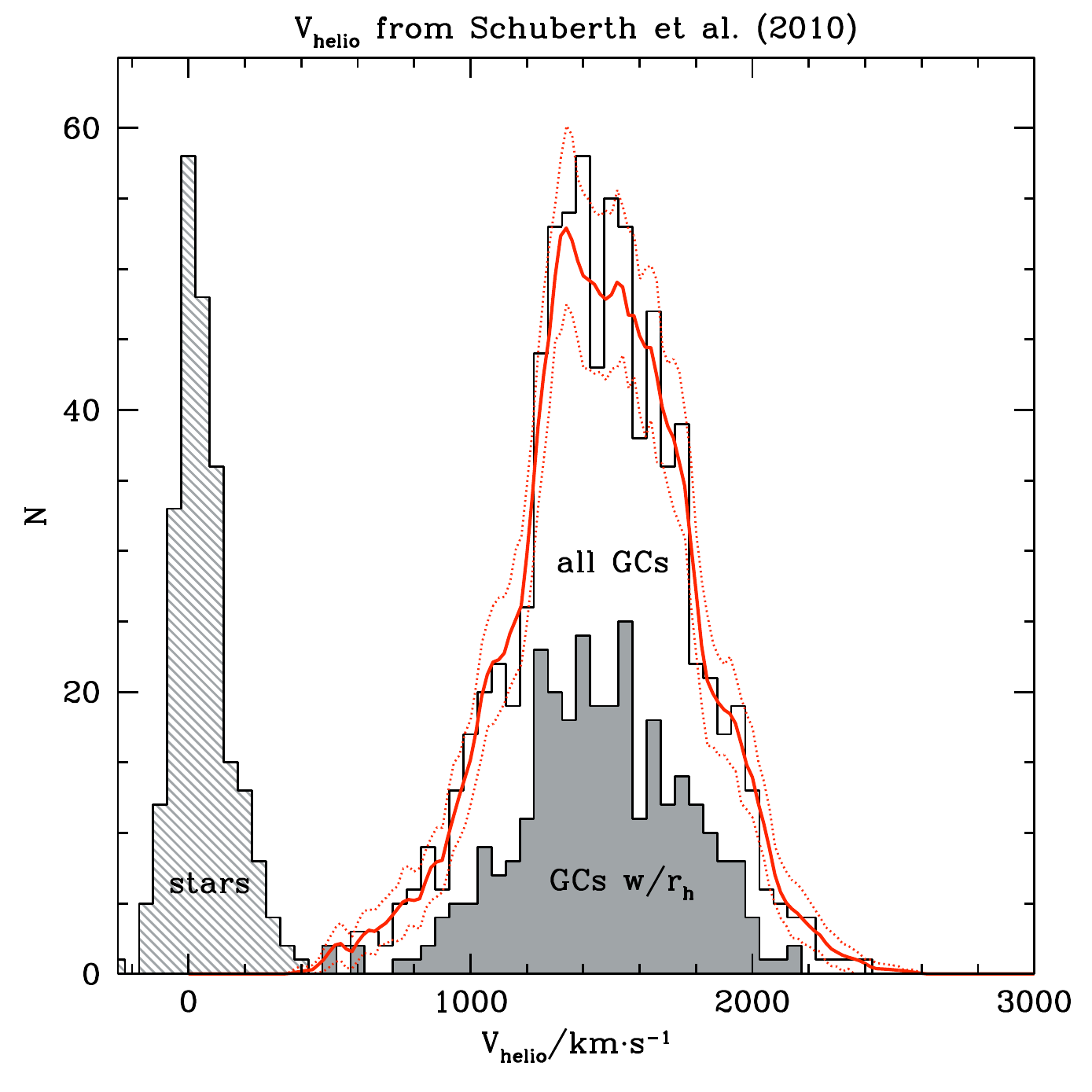}
\caption{Radial velocity distribution of objects towards the core regions of the Fornax cluster for which \cite{schuberth10} provide $v_{\rm helio}$ measurements.~The open histogram are all GCs with such  measurements and the shaded histogram shows the $v_{\rm helio}$ distribution of matched GCs for which we measured structural parameters. The hatched histogram shows the $v_{\rm helio}$ distribution of foreground stars from \cite{schuberth10}.~The solid and dotted red curves show the probability density estimates to the entire GC sample together with their 90\% confidence limits.}
\label{pdf:vrad}
\end{figure}

In the following we use radial velocity measurements from \cite{schuberth10} to define a clean GC sub-sample which is consistent with the systemic velocity and GCS velocity dispersion in the center of Fornax.~Figure~\ref{pdf:vrad} shows the distribution of heliocentric radial velocities, $v_{\rm helio}$, for foreground stars and {\it bona-fide} GCs, as well as the sub-sample of GCs for which we measured structural parameters.~We match 306 out of the 790 GCs for which \cite{schuberth10} provide $v_{\rm helio}$ values that have structural parameter measurements from our analysis.~Most of the remaining objects are at larger galactocentric radii and a small fraction has bad profile fits due to detector edge effects and confusion with very bright nearby sources.~The distribution of the matched GCs, illustrated in Figure~\ref{pdf:vrad}, shows that they representatively sample the total radial velocity distribution of the \citeauthor{schuberth10} sample.~We also match nine stars out of the 236 confirmed by \citeauthor{schuberth10}~(see the hatched histogram around $v_{\rm helio}\approx0$ km/s in Figure~\ref{pdf:vrad}) and study the distribution of structural parameters of false positives introduced by the foreground stellar population.

\subsection{GC Half-Light Radius as a Function of Luminosity}
\label{ln:rh_lum}
Before analyzing GC size variations as a function of GC color and galactocentric radius we need to make sure that potential observational biases are not influencing our result.~One such bias is a correlation between GC size and luminosity; such a correlation can introduce systematics in the size distribution function for photometrically selected samples due to changing $M/L$ ratios for stellar populations with different ages and/or metallicities. The population synthesis models of \cite{bc03} and \cite{maraston05} give roughly a factor two difference between the stellar $(M/L)_V$ ratios for 13 Gyr old stellar populations with metallicities [Z/H]~$\!=\!-1.5$ and $-0.5$ dex, which roughly correspond to the mean metallicities of the GC sub-populations in the Milky Way and other massive spiral and elliptical galaxies \citep[e.g.][]{peng06}.~For a magnitude-limited sample such a $M/L$ difference would correspond to a $\sim\!0.75$ mag offset in completeness for a uniformly old GC population.

\begin{figure}[!t]
\centering
\includegraphics[width=8.9cm]{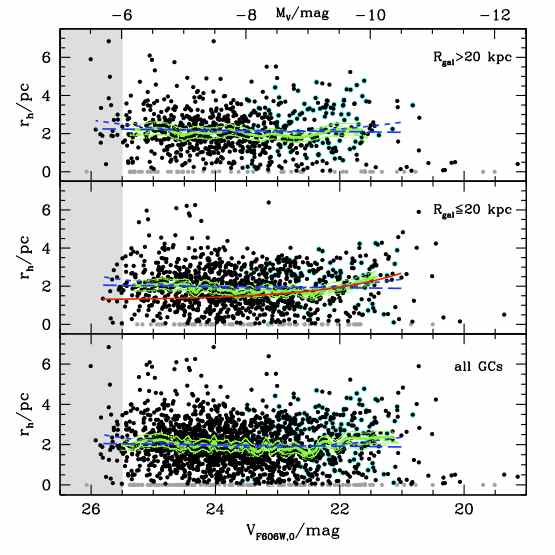}
\caption{GC half-light radius as a function integrated $V_{\rm F606W,0}$ luminosity for outer clusters ({\it top panel}), inner clusters ({\it middle panel}) and the entire GC sample ({\it bottom panel}). Black dots are resolved GC while grey dots mark unresolved objects. Solid green curves show the sliding-median trends of resolved data together with their 90\% percentile limits.~Dashed curves are the corresponding trends for all objects.~Linear and quadratic least-square fits to the resolved cluster data are shown as long-dash and short-dash lines, respectively. The shaded region at faint luminosities ($V_{\rm F606W,0}\!>\!25.5$ mag) indicates the region where the photometric pre-selection becomes significantly incomplete.~Cyan data mark {\it bona-fide} GCs confirmed by their radial velocity.}
\label{pdf:rh_intmag}
\end{figure}

We plot the GC size, i.e.~half-light radius $r_h$, versus luminosity in Figure~\ref{pdf:rh_intmag}. Running median curves with their 90\% percentile limits show that there is no indication for any significant GC size-luminosity relation for the entire GC sample.~At a constant $M/L$ ratio this corresponds to $L\!\propto\!r_h^3\rho$ and implies, therefore, that the stellar density is directly proportional to the GC size, i.e.~$\rho\!\propto\!r_h^{-3}$. Linear and quadratic least-square fits (dashed blue lines in Fig.~\ref{pdf:rh_intmag}) do not show any significant slopes for the entire sample and neither linear or higher-order fits are statistically preferred over one another.

Splitting the entire GC sample at a projected galactocentric radius of $R_{\rm gal}\!=\! 20$ kpc into an 'inner' and 'outer' sub-population, we spot a few interesting trends.~Firstly, at intermediate luminosities ($22\!\la\!V_{\rm F606W,0}\!\la\!24.0$ mag) the 'inner' sample contains fewer extended GCs with half-light radii $r_h\!\ga\!4$ pc than the 'outer' sample.~This is ruled out to be due to the varying galaxy background and/or completeness (see Section~\ref{ln:bckg}) as well as due to lower statistics in the galaxy center (as there are actually more GCs), and might be due to the preferred disruption of extended GCs in the inner regions of NGC\,1399.~The fact that we see virtually no extended GCs more massive than the GCLF turn-over at $V_{\rm F606W,0}\approx24.0$ mag indicates that disruption or tidal limitation (see Section~\ref{txt:innerouter}) may occur more frequently for low-mass GCs and that high-mass GCs are more prone to dynamical friction and orbital decay \citep[e.g.][]{lotz01}.~Secondly, we observe a weak indication for a size-luminosity relation for GC brighter than $V_{\rm F606W,0}\approx22.0$ mag, predominantly for the 'inner' sub-sample. We fit this sub-sample separately with a linear relation that yields a significant slope of $r_h\!\propto\!(-0.6\pm0.2)\, V_{\rm F606W,0}$ which is reminiscent of the transition from the GC regime without any size-luminosity relation below $M_\star\approx10^6 M_\odot$ to the size-stellar mass relation ($r_h\propto M_\star^{0.8}$) of more massive compact stellar systems such as UCDs \citep[e.g.][]{taylor10, misgeld10, misgeld11}.~This relation is indicated in the middle panel as a thin red curve and is a good representation to the general trend of the data.~We point out that the sub-sample of confirmed GCs with $v_{\rm helio}$ measurements (cyan dots in Fig.~\ref{pdf:rh_intmag}) is consistent with this trend.

Despite the fact that we detect a weak size-luminosity relation for massive GCs we stress that the majority of our sample, in particular the intermediate-luminosity to faint-end part, does not show any such relation. We are therefore safe to apply a simple magnitude cut to our data without introducing systematics in the GC size-color relation which we discuss in the following.

\subsection{GC Half-Light Radius as a Function of Color}
\label{ln:sizecol}
\begin{figure}[!t]
\centering
\includegraphics[width=8.9cm]{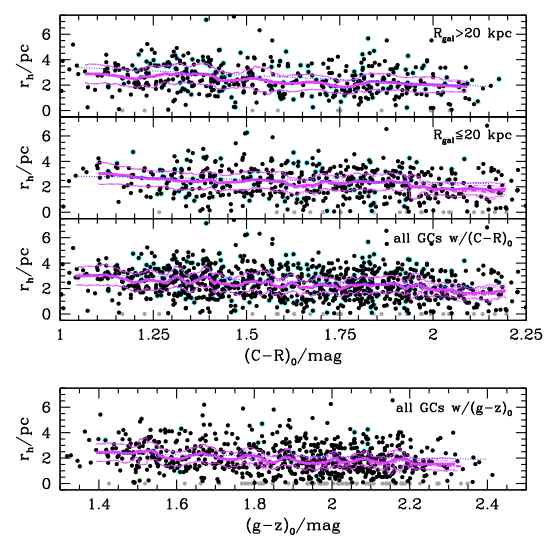}
\caption{Globular cluster half-light radius, $r_h$, as a function of their photometric color. {\it Top panel}: The bottom sub-panel shows the distribution of half-light radii for all GCs with ground-based $(C\!-\!R)_0$ color information from \cite{schuberth10}. The two upper sub-panels show the size distribution divided in projected galactocentric distance at $R_{\rm gal}\!=\!20$ kpc for the inner and outer samples. {\it Bottom panel}: GC half-light radii as a function of their $(g\!-\!z)_0$ color based on HST photometry taken from \cite{kundu08}. Solid thick and thin curves illustrate the running median and the $1\sigma$ limits, respectively, of the $r_h$ distribution in each sub-panel. We show the unresolved clusters as grey dots, which are not considered in computing the solid curves. Including those unresolved sources results in corresponding relations shown as dash-dotted curves. Cyan data illustrate {\it bona-fide} GCs confirmed by their radial-velocity.}
\label{pdf:rh_color}
\end{figure}

We add photometric color information to our GC size measurements and search the $C\!-\!R$ color database presented in \cite{schuberth10} to find 1811 sources that match our final catalog within 1\arcsec\ matching radius.~The \citeauthor{schuberth10} photometric catalog is a combination of 1) the \cite{dirsch03} Washington photometry, obtained for one central pointing with a field of view of $36\arcmin\times36\arcmin$ using the MOSAIC camera on the CTIO-Blanco 4m telescope and 2) the photometry from \cite{bassino06} which covers additional fields in the outskirts around NGC\,1399, also imaged with the MOSAIC camera.~In addition, we combine our GC size measurements with the HST photometry of \cite{kundu08} and find 1258 matches within 1\arcsec\ search radius, all of which are within the field of view of only one central HST pointing.~These two datasets have very different completeness limits and spatial resolution characteristics so that we use them only to search for differential trends in each dataset separately.

We note that based on the ACSVCS data, \cite{jordan05} have demonstrated that the mean trend of increasing half-light radius towards bluer GC colors does not strongly depend on the host galaxy $(g\!-\!z)_{\rm gal}$ color, except for the very bluest galaxies with $(g\!-\!z)_{\rm gal}<1.52$ mag where the GC size difference appears to vanish (see more detailed discussion in Sect.~\ref{ln:discussion}). In that sense, the GC system of NGC\,1399 should be representative for most massive galaxies.

In Figure~\ref{pdf:rh_color} we show the trends of GC size, i.e.~half-light radius $r_h$, versus $C\!-\!R$ color from the MOSAIC study and the $g\!-\!z$ color from the HST central pointing and find significant trends in both colors of increasing GC sizes towards bluer GC colors.~This is a different depiction of the well-known size difference between blue and red GCs discussed in previous studies \citep[e.g.][]{kundu98}.~For both photometry samples of resolved clusters we find $r_h\!\propto\!(-0.44\pm0.15)\times(g\!-\!z)$ for the HST data and $r_h \!\propto\! (-0.78\pm0.15)\!\times\!(C\!-\!R)$ for the wide-field MOSAIC sample.~For the entire MOSAIC sample the average $r_h$ gradient corresponds to a mean size difference of $\sim\!15\%$ between the peak colors $C\!-\!R=1.3$ and 1.8 mag.~Since the MOSAIC data cover a wide field of view we determine the GC size variation as a function of color for two sub-samples split at 20 kpc in projected galactocentric distance into an 'inner' and 'outer' sample.~We find that the $r_h$ gradient is stronger for the outer sample [i.e.~$r_h\!\propto\!(-0.89\pm0.22)\times(C\!-\!R)$] compared to the inner variation [i.e.~$r_h \propto (-0.65\pm0.22)\times(C\!-\!R)]$, which corresponds to a physical size variation of $\sim\!12\%$ and $\sim\!17\%$, respectively. If we use only radial-velocity confirmed GCs we obtain a much more significant $r_h$ change, namely $r_h\!\propto\!(-0.52\!\pm\!0.50)\!\times\!(C\!-\!R)$ for the inner and $r_h\!\propto\!(-1.36\!\pm\!0.45)\!\times\!(C\!-\!R)$ for the outer MOSAIC sample.~This corresponds to a physical difference of $\sim\!10\%$ and $\sim\!23\%$, respectively.~The overlap between the radial-velocity information and the HST photometry sample from \citeauthor{kundu08} is too small to derive any robust $r_h$ gradient values.~However, for illustration purposes we mark all {\it bona-fide} GCs confirmed by their $v_{\rm helio}$ as cyan dots in Figure~\ref{pdf:rh_color} and find no significant differences in their GC size-color distributions down to the limiting magnitude of $V\!\approx\!23.5$ mag, which marks the typical limit of spectroscopic studies.

\subsection{GC Half-Light Radius as a Function of Projected Galactocentric Radius}
\label{ln:sizediff}

\begin{figure*}[!t]
\centering
\includegraphics[width=14.5cm]{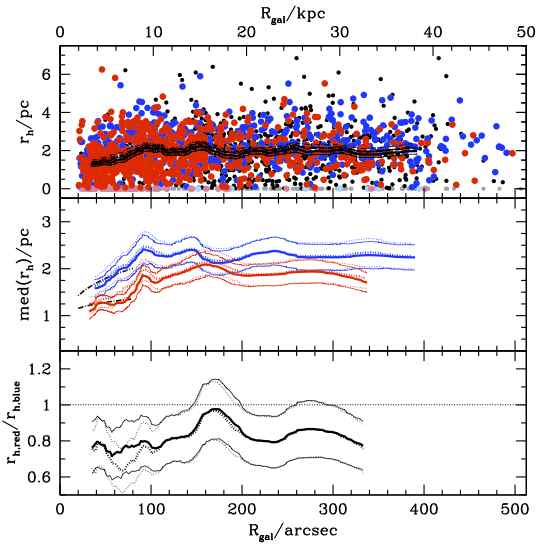}
\caption{Globular cluster half-light radius, $r_h$, as a function of projected galactocentric distance, $R_{\rm gal}$. {\it Top panel:} The plot shows all measurements for individual GCs. Blue and red dots are GCs that were classified as members of the blue and red GC sub-population by their $(C\!-\!R)$ or $(g\!-\!z)$ colors (see Table~\ref{tab:col_sel}). Black dots show GCs with structural parameter measurements that either have no ground-based colors or fall outside the fields of the corresponding studies that provide such colors (see text for details).~The solid curves show the running median with 1-$\sigma$ error of the mean margins for the entire GC sample.~The dot-dashed curves are the corresponding relations excluding unresolved objects, i.e.~$r_h\!>\!0$ pc.~{\it Middle panel:} The plot shows running-median relations for blue and red GC sub-populations illustrated as blue and red curves, respectively. The dotted relations exclude unresolved objects.~Thin curves show the 1-$\sigma$ error of the mean.~Black dash-dotted curves indicate the GC size-$R_{\rm gal}$ relations for blue and red GCs in M87 as derived by \cite{madrid09}.~{\it Bottom panel:} The ratio of the median half-light radii between red and blue GCs as function of projected galactocentric distance with the corresponding 1-$\sigma$ uncertainties and the same relations excluding unresolved objects as dotted curves.~Note that the bottom abscissa and those in between the panels show the galactocentric radius in arcseconds, while the top abscissa indicates the physical scale in kpc assuming a Fornax distance of $20.13$ Mpc.}
\label{pdf:rh_galrad}
\end{figure*}

Thanks to the wide field coverage of our ACS mosaic we are now in the position of determining the change of the {\it classic} size difference between blue and red GCs as a function of projected galactocentric radius, $R_{\rm gal}$, in much greater detail.~To begin with, we use the photometric parameters summarized in Table~\ref{tab:col_sel} to define the blue and red GC sub-sample.~The top panel of Figure~\ref{pdf:rh_galrad} shows the corresponding GC size versus~$R_{\rm gal}$ distribution for all GC candidates.~Taking the entire GC sample for which structural parameters were measured and calibrated (see Section~\ref{ln:sp}), we observe several interesting regimes with constant and gradually changing GC sizes.~Firstly, GCs in the inner $\sim\!10$ kpc become on average larger as a function of $R_{\rm gal}$, while GCs at larger galactocentric distances ($\ga\!10$ kpc) show no significant GC size-$R_{\rm gal}$ relation.~This is illustrated by the black curves which depict the sliding median together with error-of-the-mean margins.~Secondly, plotting the median size trends for the blue and red GC sub-population separately (middle panel of Fig.~\ref{pdf:rh_galrad}) reveals the well known GC size difference of $\sim\!20\%$ in the central parts of NGC\,1399, i.e.~$R_{\rm gal}\la10$ kpc \citep[e.g.][]{kundu98, jordan05}.~Except for the range $R_{\rm gal}\!\approx\!14\!-\!20$ kpc, this size difference prevails at large galactocentric distances out to $\sim\!30\!-\!40$ kpc.~The bottom panel shows the ratio of the median GC sizes for blue and red clusters in the sense ${\rm med}(r_{\rm h, red})/{\rm med}(r_{\rm h, blue})$.~This mean ratio for the whole $R_{\rm gal}$ range is $0.82\pm0.11$.~{\it The existence of a GC size difference at large $R_{\rm gal}$ is direct evidence that this difference cannot be solely due to a projection effect as suggested by \cite{larsen03}. Instead, it has to have its origin in at least one other internal or external parameter that determines the GC size and/or its evolution.~The simulations of \cite{sippel12} suggest that this size difference is mainly due to GC internal evolution related to the impact of metallicity effects on stellar evolution combined with the GC dynamical evolution under the influence of mass segregation.}

In the middle panel of Figure~\ref{pdf:rh_galrad} we show the comparison with the GC size-$R_{\rm gal}$ relations for blue and red GCs in M87 as derived by \cite{madrid09}.~Similar conclusions have been reached by \cite{paolillo11}, \cite{blom12}, and \cite{webb12b}.~Within the $R_{\rm gal}$ coverage of the single central ACS pointing that these authors have used for their analysis, the agreement between their M87 and our NGC\,1399 GC size trends is remarkably good.

\begin{figure*}[!t]
\centering
\includegraphics[width=17.5cm]{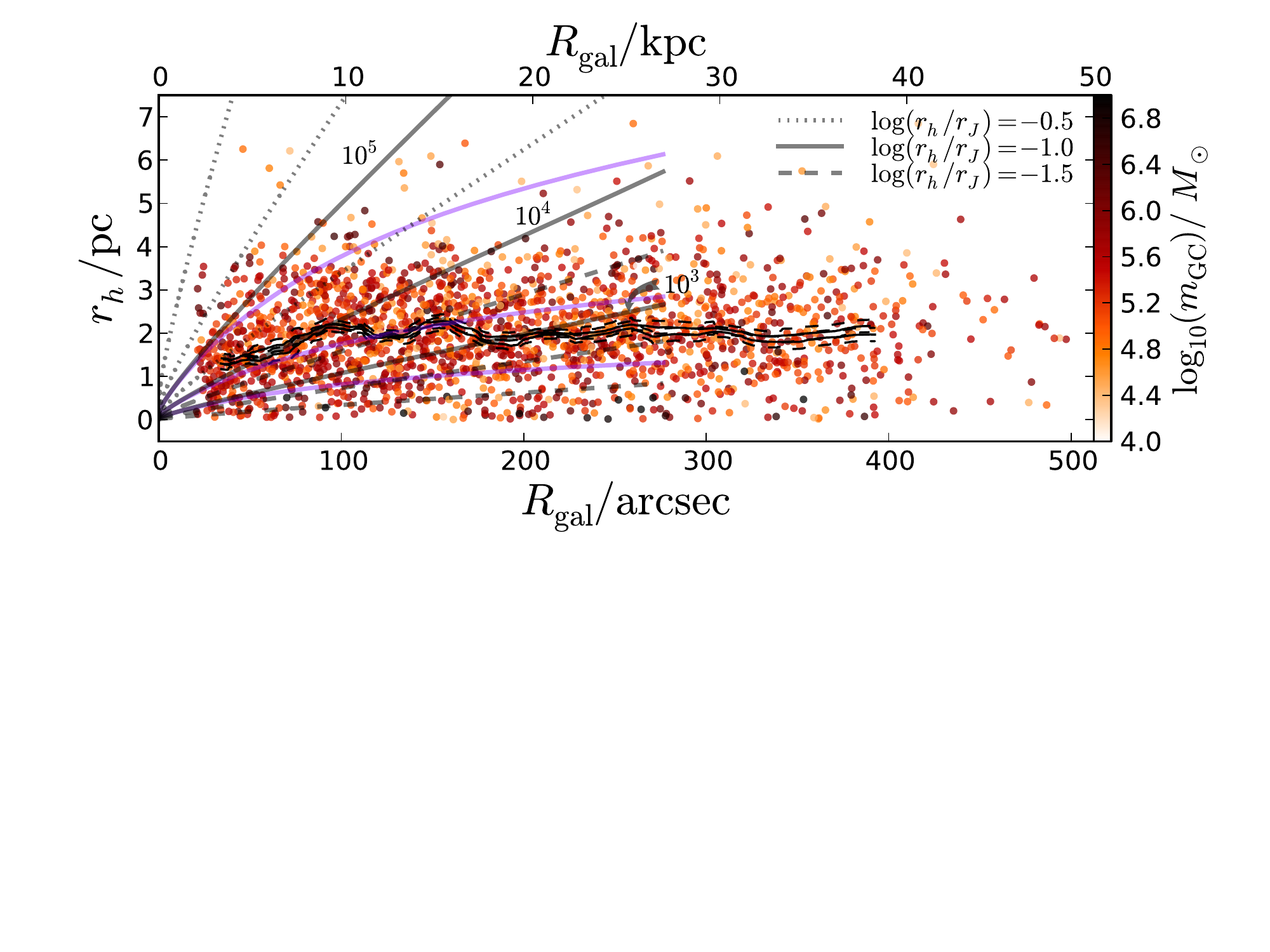}
\caption{GC half-light radius vs. galactocentric radius as in the top panel of Figure~\ref{pdf:rh_galrad}.~This time we overplot estimates of the GC half-light radius based on the derived GC Jacobi radius for GC masses, $m_{\rm GC}=10^3, 10^4$, and $10^5 M_\odot$ and  minimum, mean, and maximum ratios between the half-light and Jacobi radius, $\log(r_h/r_J)=-0.5,-1.0,-1.5$ based on the work of \cite{ernst13}. Grey shaded curves consider only the stellar mass profile of NGC\,1399, while magenta curves show the corresponding relations for the combined stellar+dark matter mass density profile (see text for details).}
\label{pdf:rh_galrad_jacobi}
\end{figure*}

\begin{figure*}[!t]
\centering
\includegraphics[width=15cm]{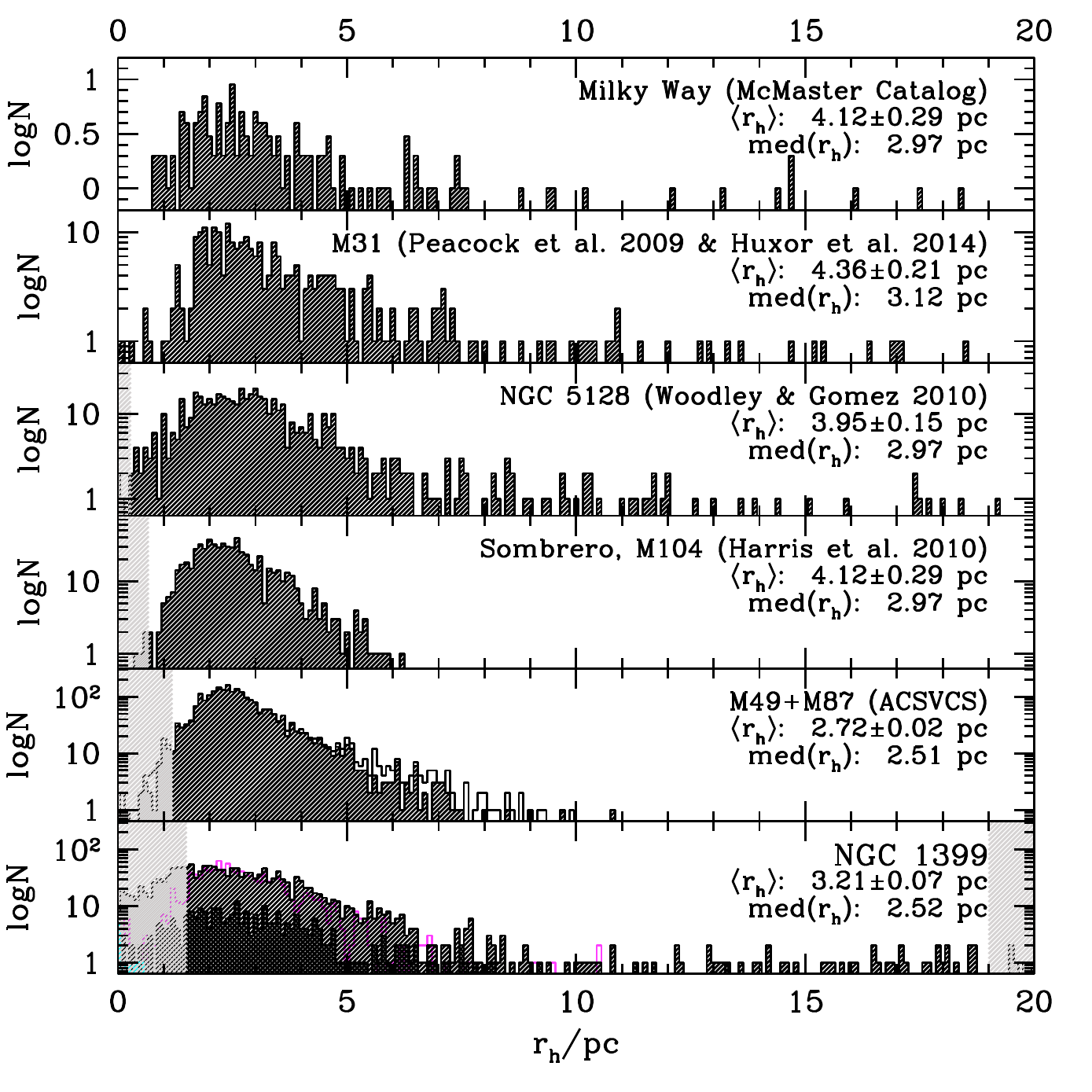}
\caption{Half-light radius distribution functions for various GC samples. The two top panels show the $r_h$ distributions for the Milky Way and M31 GC system, the measurements of which were taken from the 2010 version of the McMaster catalog \citep{harris96} as well as from \cite{peacock09} and \cite{huxor14}, respectively. The other panels below show the corresponding $r_h$-distributions for GCs in NGC~5128 \citep{woodley10}, the Sombrero galaxy \citep[M\,104,][]{harris10}, and the two brightest Virgo ellipticals M\,49 and M\,87, studied by the ACSVCS \citep{jordan09}. To illustrate the variation in selecting GC photometrically from ACSVCS data, we plot for the two Virgo galaxies the distributions for all objects with a GC-likelihood parameter of $p=0.9$ (likely genuine GCs, {\it shaded histogram}) and $p=0.2$ (GCs and objects that are less likely of GC nature, {\it open histogram}; see \citeauthor{jordan09} for details). The bottom panel shows our $r_h$ measurements for all NGC\,1399 GCs as shaded histogram and for all spectroscopically confirmed GCs as dark histogram. We also show the distribution of GC half-light radii for the center region of NGC\,1399 presented in \cite{masters10} based on data from the ACS Fornax Cluster Survey \citep[{\it magenta histogram}, see][]{jordan07}. Note that all nine confirmed foreground stars cluster are as expected unresolved objects ({\it cyan histogram}).~Assuming observations in the F606W filter we show the expected resolution limits as grey shaded regions.~The grey shaded region in the bottom panel at $r_h\!\geq\!19$ pc shows the parameter space section where the correction functions are less robustly defined (see Section~\ref{ln:sp} for details).}
\label{pdf:h_distr}
\end{figure*}

\section{Discussion}
\label{ln:discussion}

\subsection{The Inner vs.~Outer GC System of NGC\,1399}
\label{txt:innerouter}
The significant GC size-luminosity relation of the inner $10$\,kpc which disappears in the outer regions may indicate a transition in the predominance of various mechanisms at different galactocentric radii that shape the GC sizes and thus their evolution as a system.~Since the transition does not depend on GC color, i.e.~blue and red GCs show that same $r_h-R_{\rm gal}$ relation, external dynamical effects are the most probable explanation (e.g.~dynamical friction of massive GCs that quickly sink into the core regions of the inner galaxy, tidal harassment of low-mass GCs by dwarf haloes in the outer halo regions, etc.).~While detailed numerical modelling of these effects goes beyond this work, we point out that our dataset is ideal to conduct detailed analyses such as those presented in \cite{vesperini03} and \cite{webb12a,webb13}.~We note that the mean $r_h$ for all resolved sources within $20\arcsec\!<\!R_{\rm gal}\!<\!120$\arcsec\ is $1.95\pm0.06$ pc, i.e.~significantly smaller than the mean value for the entire GC system of $\langle r_h\rangle\!=\!3.21\pm0.07$ pc. 

\begin{deluxetable}{lccc}[!t]
\centering
\tabletypesize{\scriptsize}
\tablecaption{Photometric selection of blue and red GCs.\label{tab:col_sel}}
\tablehead{\colhead{} & \colhead{blue GCs} & \colhead{red GCs} & \colhead{Ref.}}
\startdata
Ground-based & $T1<23$& $T1<23$ &  (1) \\
data & $1.0\leq C\!-\!R<1.65$ & $1.65\leq C\!-\!R<2.2$ &\\\\
HST data & $z<22.5$ & $z<22.5$ & (2) \\
 & $1.3\leq g\!-\!z<1.9$ & $1.9\leq g\!-\!z<2.5$ & \\
\enddata
\tablerefs{(1): \cite{bassino06}, (2): \cite{kundu08}}
\end{deluxetable}

To test whether the stellar mass distribution in NGC\,1399 is sufficient to produce the GC $r_h\!-\!R_{\rm gal}$ trend (see Figure~\ref{pdf:rh_galrad}) we use the surface brightness profile data obtained as part of the Carnegie-Irvine Galaxy Survey \citep[CGS, see][]{ho11,li11} to compute the local {\it instantaneous} Jacobi radius of GC ($r_J$) as a function of galactocentric radius out to $\sim\!280\arcsec$ (i.e.~$\sim\!28$\,kpc) which corresponds to the maximum sampling radius of CGS. The Jacobi radius marks the point at which the gravitation forces exerted on GC member stars due to the GC potential and that of its host galaxy are equal but opposite in direction. The Jacobi radius can be expressed as
\begin{equation}
r_J=R_{\rm gal}\left(\frac{m_{\rm GC}}{2{\cal M}_{\rm gal}}\right)^{1/3}
\label{eq:jacobi}
\end{equation}
and is a robust representation of the {\it instantaneous} GC tidal radius that is induced by the surrounding tidal field \citep{innanen83,bertin08,renaud11,webb13}.

We proceed with computing the NGC\,1399 mass distribution profile using the CGS data\footnote{http://cgs.obs.carnegiescience.edu/CGS/Home.html} and the recipes outlined in \cite{bell03} to convert photometric colors into stellar mass-to-light ratios as a function of galactocentric radius \citep[see also][]{zibetti09,into13}.~We compute the ${\cal M}_\star/L_V$ profile through linear interpolation of predictions for a 13 Gyr old stellar population with variable metallicity, which is set by the measured photometric color profile of NGC\,1399, using the 2007 update of the \cite{bc03} SSP models. With the radial trend for ${\cal M}_\star/L_V$ we derive then the corresponding relation for the stellar mass 
\begin{equation}
\label{eq:massprof}
\log({\cal M}_\star)\!=\!\log({\cal M}_\star/L_V)_r\!-\!0.4[m_V(r)\!-\!D_L\!-\!M_{V,\odot}]
\end{equation}
enclosed in $R_{\rm gal}\!=\!r$, where $m_V(r)$ is the integrated magnitude derived from the galaxy surface brightness profile, $D_L$ is the luminosity distance, and $M_{V,\odot}\!=\!4.83$ mag is the absolute $V$-band magnitude of the Sun. With Equation~\ref{eq:jacobi} and the derived stellar mass profile of NGC\,1399 from Equation~\ref{eq:massprof}, we determine the {\it instantaneous} Jacobi radii for GCs with a total mass of $m_{\rm GC}\!=\!10^3, 10^4$, and $10^5 M_\odot$ using the results from \cite{baumgardt10} and \cite{ernst13} who determine the typical ratios between half-light and Jacobi radius with minimum, mean, and maximum values of $\log(r_h/r_J)\!=\!-1.5,-1.0$, and $-0.5$ for Milky Way GCs. The extremes of the $r_h/r_J$ distribution are representative of GCs that are under- and overfilling their Roche lobes, respectively. 

We also compute the GC stellar masses using the differential ${\cal M}_\star/L$ predictions from the GALEV SSP models \citep{kotulla09}, assuming uniformly old GC ages ($t_{\rm GC}=13$\,Gyr) and using the $g\!-\!z$ and $C\!-\!T1$ GC colors to account for ${\cal M}_\star/L$ variations as a function metallicity. For GCs which lack color information we adopt the median ${\cal M}_\star/L$ of the GC sample for which photometric colors are available.

We overplot the corresponding expectation trends for $r_h$ as a function of galactocentric radius in Figure~\ref{pdf:rh_galrad_jacobi} and use the color shading to parametrize GC mass.~Consistent with Figure~\ref{pdf:rh_intmag} we see no preferred GC mass scale at a given galactocentric radius. We observe that none of the curves reproduces the break at 10\,kpc of the $r_h\!-\!R_{\rm gal}$ profile and its flatness at large galactocentric radii. The stellar mass density distribution is clearly not sufficient and requires an additional mechanism to limit GC sizes at large $R_{\rm gal}$.~This could in principle be achieved by an exotic eccentricity distribution function of GC orbits, which would bring the outer clusters into the inner galaxy on preferentially radial orbits \citep[see][]{webb13}.~An alternative explanation for the observed situation could be an additional tidal limitation of GCs in the outskirts of the galaxy, which could be realized in two different ways: 

\noindent 1) by an additional mass component in the form of a dark matter density profile of the NFW type $\rho(r)\!=\!\rho_0/[(r/R_s)(1+r/R_s)^2]$ \citep{navarro96} where the total mass inside radius $R_{\rm gal}$ is given by
\begin{eqnarray}
{\cal M}_{\rm DM}(<\!R_{\rm gal})&=& 4\pi \int_0^{R_{\rm gal}}r^2 \rho (r) dr\\
    &=&4\pi \rho_0 R_s^3 \left[\ln\left(\frac{R_s+R_{\rm gal}}{R_s}\right)-\frac{R_{\rm gal}}{R_s+R_{\rm gal}}\right]\!. \nonumber
\end{eqnarray}
The resulting relations for $\rho_0\!\approx\!4\cdot10^7 M_\odot/{\rm pc}^3$, $R_s\!\approx\!130$\,kpc, and $\log(r_h/r_J)\!=\!-1.0$ are illustrated in Figure~\ref{pdf:rh_galrad_jacobi} as magenta curves and show that even in the presence of a typical dark matter halo, i.e.~considering ${\cal M}_\star\!+\!{\cal M}_{\rm DM}\!=\!{\cal M}_{\rm gal}$ in Equation~\ref{eq:jacobi}, the $r_h\!-\!R_{\rm gal}$ GC relations are still monotonically increasing, albeit not as rapidly as in the case of considering ${\cal M}_\star$ only. Hence, an additional component is required to flatten out the $r_h\!-\!R_{\rm gal}$ profiles at large galactocentric radii. 

\noindent 2) We, therefore, suggest that an increased stochastic distribution of low-mass dark matter haloes that are part of the galaxy cluster potential induce additional tidal stress on outer-halo GCs.~Such a changing mass fraction in subhaloes as a function of galactocentric radius is observed in high-resolution $\Lambda$CDM simulations \citep[e.g.][]{springel08} and would increase the ``tidal variance" in outer-halo regions, thereby truncating the GC stellar density profiles.~This may limit the GC sizes to a roughly constant value, something that shall be explored with dedicated high-resolution numerical simulations.

\subsection{Structural Parameter Distributions}
\label{ln:distribs}

\begin{deluxetable*}{lccccccccccccc}[!t]
\centering
\tabletypesize{\scriptsize}
\tablecaption{Fractions of extended GCs for various GC systems.\label{tab:radfrac}}
\tablehead{\colhead{Host Galaxy} & \colhead{$E_5$} & \colhead{$\hat{E}_5$} &\colhead{Ref.} &\colhead{Dist./Mpc} &\colhead{Ref.} & \colhead{$R_{\rm gal}$/kpc} & \colhead{$E_{5/10}$} & \colhead{$\hat{E}_{5/10}$} &\colhead{$r_e(K_s)$} &\colhead{$r_e$/kpc} & \colhead{${\cal E}_{5}$} & \colhead{$\hat{{\cal E}}_{5}$} }
\startdata
NGC~1399 			& 0.122 & 0.62 & (1) & $20.13\pm0.4$ & (7) & 51.3 & 0.061 & 0.21 & 32.9\arcsec & 3.21& 0.0480 & 0.12 \\
NGC~4486 (M87)		& 0.066 & 0.34 & (2) & $16.70\pm0.2$ & (8) & 12.3 & 0.064 & 0.22 & 41.5\arcsec & 3.36& 0.0633 & 0.16 \\
NGC~4472 (M49)		& 0.073 & 0.37 & (2) & $16.40\pm0.2$   & (8) & 11.6 & 0.072 & 0.25 & 56.1\arcsec & 4.46& 0.0714 & 0.18 \\
NGC~4594 (M104)		& 0.026 & 0.13 & (3) & $9.08\pm0.2$   & (9) & 15    & 0.024 & 0.08 & 55.3\arcsec & 2.43& 0.0160 & 0.04 \\
NGC~5128 (Cen\,A)		& 0.170 & 0.86 & (4) & $3.84\pm0.35$ &(10)& 20    & 0.179 & 0.61 & 82.6\arcsec & 1.54& 0.2127 & 0.54 \\
NGC~224   (M31)		& 0.241 & 1.22 & (5) & $0.779\pm0.05$&(11)& 160  & 0.132 & 0.45 & 443.2\arcsec & 1.67& 0.1316 & 0.33 \\
Milky Way            		& 0.197 &$\equiv1$&(6)& $\dots$&$\dots$&120&0.292&$\equiv1$&$\dots$&2.50&0.3974&$\equiv1$\\
\enddata
\tablecomments{$R_{\rm gal}$ is the maximum sampling radius of the corresponding dataset in kpc.~$E_5$ and $\hat{E}_5$ are the values defined in Equations~\ref{eq:S5} and \ref{eq:S5hat}, while the corresponding values for the GC samples restricted to $R_{\rm gal}\!\leq\!10$\,kpc are given as $E_{5/10}$ and $\hat{E}_{5/10}$ and those within 2.5 effective radii as ${\cal E}_{5}$ and $\hat{{\cal E}}_{5}$, respectively.~$K_s$-band effective radius measurements, $r_e(K_s)$, are from 2MASS and were obtained from the NASA/IPAC Infrared Science Archive.~For the Milky Way, the corresponding value was adopted based on the predictions of the Besan\c{c}on Galactic stellar population synthesis model \citep{robin03}.}
\tablerefs{For the GC populations, (1): this work, (2): ACSVCS, see \cite{jordan09}, (3): \cite{harris10}, (4): \cite{woodley10}, (5): \cite{peacock09} and \cite{huxor14}, (6): McMaster catalog, 2010 update of \cite{harris96}. For the distance measurements, (7): \cite{dunn06}, (8): \cite{mei07}, (9): \cite{jensen03}, (10): \cite{harrisG10}, (11): \cite{conn12}.}
\end{deluxetable*}

We show the $r_h$ distribution of NGC\,1399 GCs in Figure~\ref{pdf:h_distr} together with corresponding measurements for Milky Way and M31 GCs, taken from the McMaster catalog \citep[2010 update of][]{harris96} as well as \cite{peacock09} and \cite{huxor14}, respectively.~In addition, we compare our half-light radius measurements to the $r_h$ distributions of GCs in NGC~5128 \citep{woodley10}, the Sombrero galaxy \citep[M104,][]{harris10}, and the two brightest Virgo ellipticals M49 and M87 which were studied by the ACSVCS \citep[for details see][]{jordan09}.

The bottom panel of Figure~\ref{pdf:h_distr} shows the entire sample of NGC\,1399 GCs together with the distribution of radial-velocity confirmed GCs (dark histogram) and stars (cyan histogram).~It is important to note that all spectroscopically confirmed foreground stars concentrate around $r_h\!\approx\!0$ pc, where unresolved objects are generally expected.~We provide mean and median values of each $r_h$ distribution in each panel of Figure~\ref{pdf:h_distr} and point out that there is a trend of decreasing $r_h$ with increasing host galaxy luminosity \citep{masters10} in which NGC\,1399 and its central GC system fit right in. Such a trend generally supports the notion that the host environment has an impact on the GC $r_h\!-\!R_{\rm gal}$ relation (see discussion above), and will depend on the sampled $R_{\rm gal}$ range.

We also compare our sample to the measurements of \cite{masters10} who derived GC half-light radii for the central regions in NGC\,1399 from the ACS Fornax Cluster Survey data \citep{jordan07} which, similar to the ACSVCS in Virgo, sampled massive early-type galaxies in Fornax with one central HST/ACS pointing. Both our and the ACSFCS distributions show very similar shapes and drop-offs from $\sim\!1.5$ pc up to about 5 pc, beyond which our sample starts to include many more extended GCs. This is mainly due to the nine times larger field of view of our data and we point out that many of these extended sources are radial-velocity confirmed {\it bona-fide} GCs at large galactocentric radii.

The comparison with other GC systems in the upper panels of Figure~\ref{pdf:h_distr} shows that all half-light radius distributions have very similar shapes featuring a relatively steep increase in GC number density at low $r_h$ values, with a peak somewhere in the range of $2\!-\!3$ pc, and a shallower decline towards more extended objects.~This distribution is dependent on the sampling of the GC luminosity function, galactocentric radius, as well as the amount of contamination, the measurement errors, and the GC selection criteria \citep[e.g.][]{brescia12}.~It is hard to compare the unresolved parts at $r_h\!\la\!1$ pc for galaxies further away than Sombrero (NGC\,4594 at $D\!\approx\!9$ Mpc) due to the resolution limit of HST (see shaded regions in Figure~\ref{pdf:h_distr}).~Despite this limitation there is ample information and some intriguing aspects of the GC size distributions for sources with $r_h\!\ga\!1.5$ pc.~In NGC\,1399, these extended clusters predominantly reside at projected galactocentric radii, $R_{\rm gal}$, larger than 10 kpc (see Figure~\ref{pdf:rh_galrad}).~Since the observed GC populations in M49, M87, and M104 are all inside this radius (see Table~\ref{tab:radfrac}), we find a very small population of similarly extended GCs in the corresponding samples.~This is, of course, an observational bias considering our and the earlier results by \cite{vdB91} and \cite{larsen03} who found correlations of the type $r_h\!\propto\!R_{\rm gal}^{n}$ with $n\!<\!1$, and \cite{jordan05} who suggested an analytic expression that approximates the $r_h$ distribution for the inner GC systems in Virgo ellipticals.

Having sampled a significant population of GCs to large galactocentric radii in NGC\,1399 in combination with similar results for less rich GC system (see Figure~\ref{pdf:h_distr}), we, therefore, suggest that all GC systems are comprised of two components of clusters: one standard GC population with a size distribution resembling the typical GC half-light radius of $2\!-\!3$ pc and a second, less rich component of more extended GCs that are predominantly found at larger galactocentric radii. Alternatively, there might be a combination of mechanisms (explored further below) that act on just one GC population, but their effects manifest themselves at different radii, so that the extended GCs are only observed at large $R_{\rm gal}$.

To quantify the fraction of extended GCs in a GC system, we define the number ratio of GCs with sizes larger than 5 pc relative to the total GC population,
\begin{equation}
\label{eq:S5}
E_5=N_{\rm GC}(r_h\!\geq\!5 {\rm pc})/N_{\rm GC}({\rm all}), 
\end{equation}
and normalize this value to the Galactic GC system, i.e.
\begin{equation}
\label{eq:S5hat}
\hat{E}_5=\frac{ N_{\rm GC}(r_h\geq5 {\rm pc})}{N_{\rm GC}({\rm all})} 
\left(\frac{N_{\rm GC}(r_h\geq5 {\rm pc})_{\rm MW}}{N_{\rm GC}
({\rm all})_{\rm MW}}\right)^{-1}.
\end{equation}
The results for all GC systems are summarized in Table~\ref{tab:radfrac} for the galactocentric sampling ranges of the corresponding dataset, which vary by about an order of magnitude. 

In order to representatively compare the GC samples we, therefore, restrict each dataset to within $R_{\rm gal}\!\leq\!10$\,kpc (about the maximum homogeneous sampling radius of the samples) as well as 2.5 effective radii of the host galaxy's diffuse light (set by the maximum radial sampling of each dataset), measured in the near-infrared $K_s$ filter.~We summarize the corresponding values as $E_{5/10}$ and $\hat{E}_{5/10}$ for $R_{\rm gal}\!\leq\!10$\,kpc as well as ${\cal E}_{5}$ and $\hat{{\cal E}}_{5}$ for $R_{\rm gal}\!\leq\!2.5\,r_e$ in Table~\ref{tab:radfrac}.

We find a clear dichotomy in the ${\cal E}_{5}$ (and $E_{5/10}$) between late-type and early-type galaxies.~While the three giant ellipticals NGC\,1399, M87 and M49 as well as M104 show ${\cal E}_{5}$ values clearly below 10\%, the two late-type spirals, i.e.~M31 and the Milky Way, as well as NGC\,5128 stand out with significantly higher ${\cal E}_{5}$ values, clearly above $\sim\!10\%$.~We attribute this result to differences in the tidal environment properties throughout the dynamical evolution and merging history of these galaxies.~Giant ellipticals experience in general a more violent evolution than spirals. It is unclear yet, how these numbers compare to other GC systems, but the fact that ${\cal E}_{5}$ values of NGC\,1399 and the two Virgo giant ellipticals M87 and M49 are remarkably similar, hints at physical processes acting that are acting in a similar way on the size evolution of their GC systems.~This includes the somewhat surprising result for the Sombrero galaxy's GC system with a similar ${\cal E}_{5}$ value as the giant ellipticals. Higher ${\cal E}_{5}$ values for the Milky Way and M31 might be the result of the dynamically more benign tidal field around such distant GCs and/or the younger, i.e.~less evolved, nature of NGC\,5128, a recent merger remnant, and its GC system. How these numbers will play out for the GC systems in other Virgo cluster galaxies will be shown by the {\it Next Generation Virgo Cluster Survey} (NGVS) which achieves a spatial resolution of $\sim\!5$\,pc for the entire Virgo galaxy cluster out to its virial radius \citep{ferrarese12, munoz14}. At least then it will be clear whether late-type galaxies have a systematically larger population of extended GCs than early-type galaxies, which host GC systems with a relatively smaller population of extended GCs.

Of course, we expect a complex interplay between the formation paths of the compact and extended GCs. In fact, we expect multiple components in the GC size distribution depending on the star cluster formation history and the evolution of the host galaxy. However, in a simplistic picture we speculate that while the primary component GCs (i.e.~compact GCs) are likely massive and old, and formed {\it in-situ}, the nature of secondary component GCs (i.e.~extended GCs) is likely the result of a combination of populations of 1) dissolving star clusters triggered by recent formation of younger, low-mass GCs combined with increased tidal stress, e.g.~in central regions of galaxy clusters or merger remnants \citep{gieles11, goudfrooij12}, 2) the accretion of more extended GCs from satellite galaxies which formed and survived in a more benign tidal environment \citep[see also][]{georgiev09, dacosta09, smith13}, and/or 3) disrupting cores of stripped dwarf galaxy nuclei \citep[e.g.][]{oh00, bekki03, pfeffer13}.~The corresponding detailed analysis of this scenario is the focus of a forthcoming paper.

\subsection{Kinematic Properties of Compact and Extended GCs} 

\begin{figure}[!t]
\centering
\includegraphics[width=8.9cm]{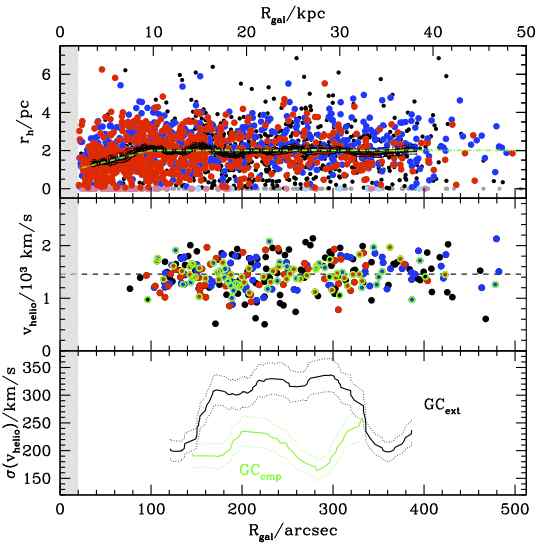}
\caption{({\it Top panel}): The plot shows all measurements for individual GCs as in Figure~\ref{pdf:rh_galrad} with the green dash-dotted curve approximating the running median relation as defined in Equation~\ref{eqn:split}.~We use this empirical separation to define formally compact and extended GCs.~({\it Middle panel}): The radial velocity of each GC matched with the \cite{schuberth10} sample as a function of galactocentric radius.~The mean radial velocity of the sample is shown as horizontal dashed line.~The symbols are parametrized by the GC color (see Table~\ref{tab:col_sel}) and split into blue and red GCs (shown in corresponding colors) and those without color information (shown as black dots). Green circles indicate GCs that have sizes formally more compact than relation shown as dash-dotted relation in the top panel.~({\it Bottom panel}):~Sliding-median relations of the line-of-sight radial velocity dispersion as a function of galactocentric radius for compact (GC$_{\rm cmp}$, {\it green}) and extended GCs (GC$_{\rm ext}$, {\it black}) with their corresponding 90\% confidence limits shown as dotted curves.}
\label{pdf:rh_galrad_losvd}
\end{figure}

The availability of matched GC size and radial velocity measurements in NGC\,1399 allows us to investigate correlations between these two parameters. The radial velocities of all GCs with size measurements do not correlate in any statistically significant way as a function of projected galactocentric radius (Figure~\ref{pdf:rh_galrad_losvd}).~We measure a total systemic heliocentric radial velocity of the entire sample as $\langle v_{\rm helio}\rangle\!=\!1456\!\pm\!17$ km s$^{-1}$ with a line-of-sight velocity dispersion of $\sigma\!=\!295$ km s$^{-1}$ in good agreement with \cite{schuberth10}.~This is also in good agreement and consistent with previous measurements of the diffuse light, i.e.~$\langle v_{\rm helio}\rangle\!=\!1425\!\pm\!4$ km s$^{-1}$ and $\sigma_0\!=\!353\pm19$ km s$^{-1}$ \citep{graham98}, respectively.

Next, we divide our sample in a population of compact (GC$_{\rm cmp}$) and extended GCs (GC$_{\rm ext}$) using as division the relation illustrated as green dash-dotted line in the top panel of Figure~\ref{pdf:rh_galrad_losvd}, which is approximating the running-median $r_h$ curve of the entire sample (black curves). This linear separation can be numerically expressed as
\begin{equation}
r_h [{\rm pc}] = \left\{ 
  \begin{array}{l l}
           0.012\,R_{\rm gal} + 0.8 & \quad \text{if $R_{\rm gal}<100$\arcsec}\\
                                               2 & \quad \text{if $R_{\rm gal}\geq100$\arcsec.}
  \end{array} \right.
  \label{eqn:split}
\end{equation}
We scrutinize the GC size-$v_{\rm helio}$ relation for any correlations and find no significant slope for compact and extended GCs as a function of galactocentric radius $R_{\rm gal}$.~However, looking at the line-of-sight velocity dispersion, $\sigma$, of each of those sub-samples, we find a surprisingly clear dichotomy between compact and extended GCs in terms of their mean velocity dispersion.~While the compact GC sample exhibits $\langle\sigma_{\rm cmp}\rangle\!=\!225\!\pm\!25$ km s$^{-1}$, we compute a much higher value for the extended sample with $\langle\sigma_{\rm ext}\rangle\!=\!317\!\pm\!21$ km s$^{-1}$.~This is consistent with the $\sigma$ differences found by \cite{schuberth10} between the blue and red GC sub-population in NGC\,1399 at a similar range in galactocentric radius. 

Plotting the sliding median of the $\sigma\!-\!R_{\rm gal}$ relation (bottom panel of Figure~\ref{pdf:rh_galrad_losvd}) reveals that this difference is most pronounced in the range $50\arcsec\!\la\! R_{\rm gal}\!\la\!320\arcsec$, which roughly corresponds to the physical range of $15\!\la\! R_{\rm gal}\!\la\!32$\,kpc.~Outside this range, the difference seem to disappear, but we lack sample statistics to make definitive conclusions, and defer a more detailed analysis of this surprising result to a future study, when more comprehensive radial velocity samples become available.~Here we just note that given the scatter of the rather weak correlation between GC size and color (see Section~\ref{ln:sizecol}), the significantly lower velocity dispersion of more compact (i.e.~red) GCs compared to their more extended counterparts (i.e.~blue GCs) appears to be the astrophysically stronger relation, which likely has its origin in the stronger influence of {\it external} tidal truncation effects compared to {\it internal} mechanisms that govern the GC size.~This is also consistent with our result of the flatter GC size-color relation for the inner vs.~outer GC sample (see Figure~\ref{pdf:rh_color}).~These findings indicate the preferential influence of {\it external} dynamical effects damping the size difference between red and blue GCs which is predominantly driven by {\it internal} evolution of their constituent stellar populations and is likely a corollary of the GC size-dynamics correlation. Future GC radial velocity samples of the inner GC system in NGC\,1399 will shed light on how the GC orbit distribution function influences these relations.

\acknowledgments
\noindent
{\it Acknowledgments} -- Support for HST program GO-10129 was provided by NASA through a grant from the Space Telescope Science Institute, which is operated by the Association of Universities for Research in Astronomy, Incorporated, under NASA contract NAS5-26555.~This research was supported by FONDECYT Regular Project Grant No.~1121005 and BASAL Center for Astrophysics and Associated Technologies (PFB-06).~THP is thankful for the hospitality and support during his visits at the University of Napoli Federico II where parts of this work were completed; he also gratefully acknowledges support in form of a Plaskett Research Fellowship from the National Research Council of Canada.~MP acknowledges financial support from the FARO\,2011 project of the University of Napoli Federico II.~We are grateful to Anton Koekemoer and Andy Fruchter for their technical support and useful discussions on the MultiDrizzle code and to Chien Y.~Peng for his help with the implementation and testing of the modified GALFIT routine.~We thank Tom Richtler and Ylva Schuberth for providing their radial velocity measurements ahead of publication, as well as Luis Ho and Zhao-Yu Li for kindly making available to us their latest NGC\,1399 surface brightness profile measurements from the Carnegie-Irvine Galaxy Survey, again prior to publication.~Avon Huxor has very kindly supplied M31 GC data prior to publication.~We are grateful to the referee, Bill Harris, for providing a thoughtful and constructive report that helped improve the presentation of the results.~We thank Jeremy Webb, Mark Gieles, Andres Jord{\'a}n, Eric Peng, Chunyan Jiang, Stephen Zepf, and Arunav Kundu for valuable discussions and providing data in electronic format.~Some of the data presented in this paper were obtained from the Multimission Archive at the Space Telescope Science Institute (MAST). STScI is operated by the Association of Universities for Research in Astronomy, Inc., under NASA contract NAS5-26555.~Support for MAST for non-HST data is provided by the NASA Office of Space Science via grant NNX09AF08G and by other grants and contracts.~This research has made use of the NASA/ IPAC Infrared Science Archive, which is operated by the Jet Propulsion Laboratory, California Institute of Technology, under contract with the National Aeronautics and Space Administration.~Figure~\ref{pdf:full} and \ref{pdf:example} were created with the help of the ESA/ESO/NASA Photoshop FITS Liberator.~This research has made use of NASA's Astrophysics Data System.

\vspace{0.2cm}
\noindent
Facilities: \facility{HST(ACS)}.

\clearpage


\begin{thebibliography}{}
\bibitem[Anderson \& King(2000)]{anderson00} Anderson, J., \& King, I.~R.\ 2000, \pasp, 112, 1360
\bibitem[Anderson \& King(2006)]{anderson06} Anderson, J., \& King, I.\ 2006, Instrument Science Report ACS 2006-001 (Baltimore: STScI), http://www.stsci.edu/hst/acs/documents/isrs/isr0601.pdf
\bibitem[Anderson(2005)]{anderson05} Anderson, J.\ 2005, The 2005 HST Calibration Workshop, eds. Koekemoer, A.~M., Goudfrooij, P., \& Dressel, L.~L.
\bibitem[Barmby et al.(2007)]{barmby07} Barmby, P., McLaughlin, D.~E., Harris, W.~E., Harris, G.~L.~H., \& Forbes, D.~A.\ 2007, \aj, 133, 2764
\bibitem[Bassino et al.(2006)]{bassino06} Bassino, L.~P., Faifer, F.~R., Forte, J.~C., Dirsch, B., Richtler, T., Geisler, D., \& Schuberth, Y.\ 2006, \aap, 451, 789
\bibitem[Bastian et al.(2008)]{bastianetal08} Bastian, N., Gieles, M., Goodwin, S.~P., Trancho, G., Smith, L.~J., Konstantopoulos, I., \& Efremov, Y.\ 2008, \mnras, 389, 223
\bibitem[Baumgardt \& Makino(2003)]{baumgardt03} Baumgardt, H., \& Makino, J.\ 2003, \mnras, 340, 227
\bibitem[Baumgardt et al.(2010)]{baumgardt10} Baumgardt, H., Parmentier, G., Gieles, M., \& Vesperini, E.\ 2010, \mnras, 401, 1832
\bibitem[Beckwith et al.(2006)]{beckwith06} Beckwith, S.~V.~W., et al.\ 2006, \aj, 132, 1729
\bibitem[Bekki \& Freeman(2003)]{bekki03} Bekki, K., \& Freeman, K.~C.\ 2003, \mnras, 346, L11
\bibitem[Bekki(2010)]{bekki10} Bekki, K.\ 2010, \mnras, 401, 2753
\bibitem[Bell et al.(2003)]{bell03} Bell, E.~F., McIntosh, D.~H., Katz, N., \& Weinberg, M.~D.\ 2003, \apjs, 149, 289
\bibitem[Bertin \& Arnouts(1996)]{bertin96} Bertin, E., \& Arnouts, S.\ 1996, \aaps, 117, 393
\bibitem[Bertin \& Varri(2008)]{bertin08} Bertin, G., \& Varri, A.~L.\ 2008, \apj, 689, 1005
\bibitem[Binney \& Tremaine(1987)]{binney87} Binney, J., \& Tremaine, S.\ 1987, Princeton, NJ, Princeton University Press, p. 235 
\bibitem[Blakeslee et al.(2009)]{blakeslee09} Blakeslee, J.~P., et al.\ 2009, \apj, 694, 556
\bibitem[Blom et al.(2012)]{blom12} Blom, C., Spitler, L.~R., \& Forbes, D.~A.\ 2012, \mnras, 420, 37
\bibitem[Bournaud et al.(2008)]{bournaud08} Bournaud, F., Duc, P.-A., \& Emsellem, E.\ 2008, \mnras, 389, L8
\bibitem[Brescia et al.(2012)]{brescia12} Brescia, M., Cavuoti, S., Paolillo, M., Longo, G., \& Puzia, T.\ 2012, \mnras, 421, 1155
\bibitem[Bruzual \& Charlot(2003)]{bc03} Bruzual, G., \& Charlot, S.\ 2003, \mnras, 344, 1000
\bibitem[Caon et al.(1994)]{caon94} Caon, N., Capaccioli, M., \& D'Onofrio, M.\ 1994, \aaps, 106, 199
\bibitem[Carlson \& Holtzman(2001)]{carlson01} Carlson, M.~N., \& Holtzman, J.~A.\ 2001, \pasp, 113, 1522
\bibitem[Chandar(2009)]{chandar09} Chandar, R.\ 2009, \apss, 324, 315
\bibitem[Chandar et al.(2010)]{chandar10} Chandar, R., Whitmore, B.~C., \& Fall, S.~M.\ 2010, \apj, 713, 1343
\bibitem[Chun et al.(1980)]{chun80} Chun, M.~S., Suh, Y.~R., \& Lee, Y.~B.\ 1980, Journal of Korean Astronomical Society, 13, 27
\bibitem[Conn et al.(2012)]{conn12} Conn, A.~R., Ibata, R.~A., Lewis, G.~F., et al.\ 2012, \apj, 758, 11
\bibitem[C{\^o}t{\'e} et al.(2004)]{cote04} C{\^o}t{\'e}, P., et al.\ 2004, \apjs, 153, 223
\bibitem[Crampton et al.(1985)]{crampton85} Crampton, D., Cowley, A.~P., Schade, D., \& Chayer, P.\ 1985, \apj, 288, 494
\bibitem[Da Costa(1979)]{dacosta79} Da Costa, G.~S.\ 1979, \aj, 84, 505
\bibitem[Da Costa et al.(2009)]{dacosta09} Da Costa, G.~S., Grebel, E.~K., Jerjen, H., Rejkuba, M., \& Sharina, M.~E.\ 2009, \aj, 137, 4361
\bibitem[Demers et al.(1990)]{demers90} Demers, S., Grondin, L., \& Kunkel, W.~E.\ 1990, \pasp, 102, 632
\bibitem[D'Ago et al.(2013)]{dago13} D'Ago et al.\ 2013, ApJ {\it submitted}
\bibitem[Dirsch et al.(2003)]{dirsch03} Dirsch, B., Richtler, T., Geisler, D., Forte, J.~C., Bassino, L.~P., \& Gieren, W.~P.\ 2003, \aj, 125, 1908
\bibitem[Downing(2012)]{downing12} Downing, J.~M.~B.\ 2012, arXiv:1204.5363
\bibitem[Drinkwater et al.(2000)]{drinkwater00} Drinkwater, M.~J., et al.\ 2000, \aap, 355, 900
\bibitem[Dunn \& Jerjen(2006)]{dunn06} Dunn, L.~P., \& Jerjen, H.\ 2006, \aj, 132, 1384
\bibitem[Elmegreen \& Efremov(1997)]{ee97} Elmegreen, B.~G., \& Efremov, Y.~N.\ 1997, \apj, 480, 235
\bibitem[Elmegreen \& Hunter(2010)]{elmegreen10} Elmegreen, B.~G., \& Hunter, D.~A.\ 2010, \apj, 712, 604
\bibitem[Elson et al.(1987)]{elson87rev} Elson, R., Hut, P., \& Inagaki, S.\ 1987, \araa, 25, 565
\bibitem[Elson(1991)]{elson91} Elson, R.~A.~W.\ 1991, \apjs, 76, 185
\bibitem[Elson(1992)]{elson92} Elson, R.~A.~W.\ 1992, \mnras, 256, 515
\bibitem[Elson \& Freeman(1985)]{elson85} Elson, R.~A.~W., \& Freeman, K.~C.\ 1985, \apj, 288, 521
\bibitem[Elson \& Walterbos(1988)]{elson88} Elson, R.~A., \& Walterbos, R.~A.~M.\ 1988, \apj, 333, 594
\bibitem[Elson \& Schade(1994)]{elson94} Elson, R.~A.~W., \& Schade, D.~J.\ 1994, \apj, 437, 625
\bibitem[Elson et al.(1987)]{elson87} Elson, R.~A.~W., Fall, S.~M., \& Freeman, K.~C.\ 1987, \apj, 323, 54
\bibitem[Ernst \& Just(2013)]{ernst13} Ernst, A., \& Just, A.\ 2013, \mnras, 429, 2953
\bibitem[Faifer et al.(2004)]{faifer04} Faifer, F.~R., Bassino, L.~P., Forte, J.~C., Dirsch, B., Richtler, T., \& Geisler, D.\ 2004, Boletin de la Asociacion Argentina de Astronomia La Plata Argentina, 47, 132
\bibitem[Fall et al.(2009)]{fall09} Fall, S.~M., Chandar, R., \& Whitmore, B.~C.\ 2009, \apj, 704, 453
\bibitem[Ferguson \& Sandage(1989)]{ferguson89} Ferguson, H.~C., \& Sandage, A.\ 1989, \apjl, 346, L53
\bibitem[Ferrarese et al.(2012)]{ferrarese12} Ferrarese, L., C{\^o}t{\'e}, P., Cuillandre, J.-C., et al.\ 2012, \apjs, 200, 4
\bibitem[Ford et al.(2003)]{ford03} Ford, H.~C., et al.\ 2003, \procspie, 4854, 81
\bibitem[Forte et al.(2005)]{forte05} Forte, J.~C., Faifer, F., \& Geisler, D.\ 2005, \mnras, 357, 56
\bibitem[Fruchter \& Hook(2002)]{fruchter02} Fruchter, A.~S., \& Hook, R.~N.\ 2002, \pasp, 114, 144
\bibitem[Fusi Pecci et al.(1994)]{fusipecci94} Fusi Pecci, F., et al.\ 1994, \aap, 284, 349
\bibitem[Georgiev et al.(2009a)]{georgiev09a} Georgiev, I.~Y., Puzia, T.~H., Hilker, M., \& Goudfrooij, P.\ 2009a, \mnras, 392, 879
\bibitem[Georgiev et al.(2009b)]{georgiev09} Georgiev, I.~Y., Hilker, M., Puzia, T.~H., Goudfrooij, P., \& Baumgardt, H.\ 2009b, \mnras, 396, 1075
\bibitem[Georgiev et al.(2010)]{georgiev10} Georgiev, I.~Y., Puzia, T.~H., Goudfrooij, P., \& Hilker, M.\ 2010, arXiv:1004.2039
\bibitem[Gieles \& Bastian(2008)]{gieles08} Gieles, M., \& Bastian, N.\ 2008, \aap, 482, 165
\bibitem[Gieles et al.(2011)]{gieles11} Gieles, M., Heggie, D.~C., \& Zhao, H.\ 2011, \mnras, 413, 2509
\bibitem[Gnedin \& Ostriker(1997)]{gnedin97} Gnedin, O.~Y., \& Ostriker, J.~P.\ 1997, \apj, 474, 223
\bibitem[G{\'o}mez et al.(2006)]{gomez06} G{\'o}mez, M., Geisler, D., Harris, W.~E., Richtler, T., Harris, G.~L.~H., \& Woodley, K.~A.\ 2006, \aap, 447, 877
\bibitem[G{\'o}mez \& Woodley(2007)]{gomez07} G{\'o}mez, M., \& Woodley, K.~A.\ 2007, \apjl, 670, L105
\bibitem[Goudfrooij(2012)]{goudfrooij12} Goudfrooij, P.\ 2012, \apj, 750, 140
\bibitem[Graham et al.(1998)]{graham98} Graham, A.~W., Colless, M.~M., Busarello, G., Zaggia, S., \& Longo, G.\ 1998, \aaps, 133, 325
\bibitem[Greif et al.(2012)]{greif12} Greif, T.~H., Bromm, V., Clark, P.~C., et al.\ 2012, \mnras, 424, 399
\bibitem[Griffen et al.(2010)]{griffen10} Griffen, B.~F., Drinkwater, M.~J., Thomas, P.~A., Helly, J.~C., 
\bibitem[Hack et al.(2003)]{hack03} Hack, W., Busko, I., \& Jedrzejewski, R.\ 2003, Astronomical Data Analysis Software and Systems XII, 295, 453
\bibitem[Harris(1996)]{harris96} Harris, W.~E.\ 1996, \aj, 112, 1487
\bibitem[Harris et al.(2006)]{harris06} Harris, W.~E., Harris, G.~L.~H., Barmby, P., McLaughlin, D.~E., \& Forbes, D.~A.\ 2006, \aj, 132, 2187
\bibitem[Harris(2009)]{harris09} Harris, W.~E.\ 2009, \apj, 699, 254
\bibitem[Harris et al.(2010)]{harris10} Harris, W.~E., Spitler, L.~R., Forbes, D.~A., \& Bailin, J.\ 2010, \mnras, 401, 1965
\bibitem[Harris et al.(2010)]{harrisG10} Harris, G.~L.~H., Rejkuba, M., \& Harris, W.~E.\ 2010, Publications of the Astronomical Society of Australia, 27, 457
\bibitem[Harris et al.(2013)]{harris13} Harris, W.~E., Harris, G.~L.~H., \& Alessi, M.\ 2013, \apj, 772, 82
\bibitem[Harris \& Pudritz(1994)]{harris94} Harris, W.~E., \& Pudritz, R.~E.\ 1994, \apj, 429, 177
\bibitem[Harris et al.(2002)]{harris02} Harris, W.~E., Harris, G.~L.~H., Holland, S.~T., \& McLaughlin, D.~E.\ 2002, \aj, 124, 1435
\bibitem[Hartwick(2009)]{hartwick09} Hartwick, F.~D.~A.\ 2009, \apj, 691, 1248
\bibitem[H{\'e}non(1973)]{henon73} H{\'e}non, M.\ 1973, \aap, 24, 229
\bibitem[H{\'e}non(1975)]{henon75} H{\'e}non, M.\ 1975, IAU Symp.~ 69: Dynamics of the Solar Systems, 69, 133
\bibitem[Hilker et al.(1999)]{hilker99} Hilker, M., Infante, L., Vieira, G., Kissler-Patig, M., \& Richtler, T.\ 1999, \aaps, 134, 75 
\bibitem[Ho et al.(2011)]{ho11} Ho, L.~C., Li, Z.-Y., Barth, A.~J., Seigar, M.~S., \& Peng, C.~Y.\ 2011, \apjs, 197, 21
\bibitem[Huxor et al.(2005)]{huxor05} Huxor, A.~P., Tanvir, N.~R., Irwin, M.~J., Ibata, R., Collett, J.~L., Ferguson, A.~M.~N., Bridges, T., \& Lewis, G.~F.\ 2005, \mnras, 360, 1007
\bibitem[Huxor et al.(2014)]{huxor14} Huxor, A.~P., et al.\ 2014, MNRAS {\it in preparation}
\bibitem[Illingworth \& Illingworth(1976)]{ill76} Illingworth, G., \& Illingworth, W.\ 1976, \apjs, 30, 227
\bibitem[Innanen et al.(1983)]{innanen83} Innanen, K.~A., Harris, W.~E., \& Webbink, R.~F.\ 1983, \aj, 88, 338
\bibitem[Into \& Portinari(2013)]{into13} Into, T., \& Portinari, L.\ 2013, \mnras, 430, 2715
\bibitem[Jee et al.(2007)]{jee07} Jee, M.~J., Blakeslee, J.~P., Sirianni, M., et al.\ 2007, \pasp, 119, 1403 
\bibitem[Jensen et al.(2003)]{jensen03} Jensen, J.~B., Tonry, J.~L., Barris, B.~J., et al.\ 2003, \apj, 583, 712
\bibitem[Jord{\'a}n(2004)]{jordan04} Jord{\'a}n, A.\ 2004, \apjl, 613, L117
\bibitem[Jord{\'a}n et al.(2005)]{jordan05} Jord{\'a}n, A., et al.\ 2005, \apj, 634, 1002
\bibitem[Jord{\'a}n et al.(2007)]{jordan07} Jord{\'a}n, A., et al.\ 2007, \apjs, 169, 213
\bibitem[Jord{\'a}n et al.(2009)]{jordan09} Jord{\'a}n, A., et al.\ 2009, \apjs, 180, 54
\bibitem[King(1962)]{king62} King, I.\ 1962, \aj, 67, 471
\bibitem[King(1966)]{king66} King, I.~R.\ 1966, \aj, 71, 64
\bibitem[King et al.(1968)]{king68} King, I.~R., Hedemann, E.~J., Hodge, S.~M., \& White, R.~E.\ 1968, \aj, 73, 456
\bibitem[Koekemoer et al.(2002)]{koekemoer02} Koekemoer, A.~M., Fruchter, A.~S., Hook, R.~N., \& Hack, W.\ 2002, The 2002 HST Calibration Workshop.~ eds. S. Arribas, A. Koekemoer, and B. Whitmore, p.337
\bibitem[Kontizas et al.(1982)]{kontizas82} Kontizas, M., Danezis, E., \& Kontizas, E.\ 1982, \aaps, 49, 1
\bibitem[Kotulla et al.(2009)]{kotulla09} Kotulla, R., Fritze, U., Weilbacher, P., \& Anders, P.\ 2009, \mnras, 396, 462
\bibitem[Kozhurina-Platais et al.(2007)]{vera07} Kozhurina-Platais, V., Goudfrooij, P., \& Puzia, T.~H.\ 2007, Instrument Science Report ACS 2007-04 (Baltimore: STScI), www.stsci.edu/hst/acs/documents/isrs/isr0704.pdf
\bibitem[Kravtsov \& Gnedin(2005)]{kravtsov05} Kravtsov, A.~V., \& Gnedin, O.~Y.\ 2005, \apj, 623, 650
\bibitem[Kron(1980)]{kron80} Kron, R.~G.\ 1980, \apjs, 43, 305
\bibitem[Kukarkin \& Kireeva(1979)]{kukarkin79} Kukarkin, B.~V., \& Kireeva, N.~N.\ 1979, Soviet Astronomy, 23, 261
\bibitem[Kundu(2008)]{kundu08} Kundu, A.\ 2008, \aj, 136, 1013
\bibitem[Kundu \& Whitmore(1998)]{kundu98} Kundu, A., \& Whitmore, B.~C.\ 1998, \aj, 116, 2841
\bibitem[Kundu et al.(1999)]{kundu99} Kundu, A., Whitmore, B.~C., Sparks, W.~B., Macchetto, F.~D., Zepf, S.~E., \& Ashman, K.~M.\ 1999, \apj, 513, 733
\bibitem[Kundu \& Whitmore(2001)]{kundu01} Kundu, A., \& Whitmore, B.~C.\ 2001, \aj, 121, 2950
\bibitem[Larsen(1999)]{larsen99} Larsen, S.~S.\ 1999, \aaps, 139, 393
\bibitem[Larsen et al.(2001)]{larsen01} Larsen, S.~S., Brodie, J.~P., Huchra, J.~P., Forbes, D.~A., \& Grillmair, C.~J.\ 2001, \aj, 121, 2974
\bibitem[Larsen \& Brodie(2003)]{larsen03} Larsen, S.~S., \& Brodie, J.~P.\ 2003, \apj, 593, 340
\bibitem[Li et al.(2005)]{li05} Li, Y., Mac Low, M.-M., \& Klessen, R.~S.\ 2005, \apj, 626, 823
\bibitem[Li et al.(2011)]{li11} Li, Z.-Y., Ho, L.~C., Barth, A.~J., \& Peng, C.~Y.\ 2011, \apjs, 197, 22
\bibitem[Lotz et al.(2001)]{lotz01} Lotz, J.~M., Telford, R., Ferguson, H.~C., et al.\ 2001, \apj, 552, 572
\bibitem[Madrid et al.(2009)]{madrid09} Madrid, J.~P., Harris, W.~E., Blakeslee, J.~P., \& G{\'o}mez, M.\ 2009, \apj, 705, 237
\bibitem[Mapelli \& Bressan(2013)]{mapelli13} Mapelli, M., \& Bressan, A.\ 2013, \mnras, 430, 3120
\bibitem[Maraston(2005)]{maraston05} Maraston, C.\ 2005, \mnras, 362, 799
\bibitem[Mar{\'{\i}}n-Franch et al.(2009)]{marinfranch09} Mar{\'{\i}}n-Franch, A., et al.\ 2009, \apj, 694, 1498
\bibitem[Masters et al.(2010)]{masters10} Masters, K.~L., Jord{\'a}n, A., C{\^o}t{\'e}, P., et al.\ 2010, \apj, 715, 1419
\bibitem[McLaughlin \& van der Marel(2005)]{mclaughlin05} McLaughlin, D.~E., \& van der Marel, R.~P.\ 2005, \apjs, 161, 304
\bibitem[McLaughlin et al.(2008)]{mclaughlin08} McLaughlin, D.~E., Barmby, P., Harris, W.~E., Forbes, D.~A., \& Harris, G.~L.~H.\ 2008, \mnras, 384, 563
\bibitem[Mei et al.(2007)]{mei07} Mei, S., Blakeslee, J.~P., C{\^o}t{\'e}, P., et al.\ 2007, \apj, 655, 144
\bibitem[Misgeld et al.(2010)]{misgeld10} Misgeld, I., Hilker, M., \& Mieske, S.\ 2010, arXiv:1010.3138
\bibitem[Misgeld \& Hilker(2011)]{misgeld11} Misgeld, I., \& Hilker, M.\ 2011, \mnras, 414, 3699
\bibitem[Monet et al.(2003)]{monet03} Monet, D.~G., et al.\ 2003, \aj, 125, 984 
\bibitem[Mu{\~n}oz et al.(2014)]{munoz14} Mu{\~n}oz, R.~P., Puzia, T.~H., Lan{\c c}on, A., et al.\ 2014, \apjs, 210, 4
\bibitem[Murphy et al.(1990)]{murphy90} Murphy, B.~W., Cohn, H.~N., \& Hut, P.\ 1990, \mnras, 245, 335
\bibitem[Murray(2009)]{murray09} Murray, N.\ 2009, \apj, 691, 946
\bibitem[Murray \& Lin(1992)]{murray92} Murray, S.~D., \& Lin, D.~N.~C.\ 1992, \apj, 400, 265
\bibitem[Navarro et al.(1996)]{navarro96} Navarro, J.~F., Frenk, C.~S., \& White, S.~D.~M.\ 1996, \apj, 462, 563
\bibitem[Oh \& Lin(2000)]{oh00} Oh, K.~S., \& Lin, D.~N.~C.\ 2000, \apj, 543, 620
\bibitem[Paolillo et al.(2011)]{paolillo11} Paolillo, M., Puzia, T.~H., Goudfrooij, P., et al.\ 2011, \apj, 736, 90
\bibitem[Peacock et al.(2009)]{peacock09} Peacock, M.~B., Maccarone, T.~J., Waters, C.~Z., Kundu, A., Zepf, S.~E., Knigge, C., \& Zurek, D.~R.\ 2009, \mnras, 392, L55
\bibitem[Peng et al.(2010)]{cpeng10} Peng, C.~Y., Ho, L.~C., Impey, C.~D., \& Rix, H.-W.\ 2010, \aj, 139, 2097
\bibitem[Peng et al.(2009)]{epeng09} Peng, E.~W., et al.\ 2009, \apj, 703, 42
\bibitem[Peng et al.(2006)]{peng06} Peng, E.~W., Jord{\'a}n, A., C{\^o}t{\'e}, P., et al.\ 2006, \apj, 639, 95
\bibitem[Pfeffer \& Baumgardt(2013)]{pfeffer13} Pfeffer, J., \& Baumgardt, H.\ 2013, \mnras, 433, 1997
\bibitem[Puzia et al.(1999)]{puzia99} Puzia, T.~H., Kissler-Patig, M., Brodie, J.~P., \& Huchra, J.~P.\ 1999, \aj, 118, 2734
\bibitem[Puzia et al.(2000)]{puzia00} Puzia, T.~H., Kissler-Patig, M., Brodie, J.~P., \& Schroder, L.~L.\ 2000, \aj, 120, 777
\bibitem[Puzia et al.(2002)]{puzia02} Puzia, T.~H., Zepf, S.~E., Kissler-Patig, M., et al.\ 2002, \aap, 391, 453
\bibitem[Renaud et al.(2011)]{renaud11} Renaud, F., Gieles, M., \& Boily, C.~M.\ 2011, \mnras, 418, 759
\bibitem[Rhode \& Zepf(2001)]{rhode01} Rhode, K.~L., \& Zepf, S.~E.\ 2001, \aj, 121, 210
\bibitem[Rhode \& Zepf(2004)]{rhode04} Rhode, K.~L., \& Zepf, S.~E.\ 2004, \aj, 127, 302
\bibitem[Rhode et al.(2007)]{rhode07} Rhode, K.~L., Zepf, S.~E., Kundu, A., \& Larner, A.~N.\ 2007, \aj, 134, 1403
\bibitem[Rhodes et al.(2007)]{rhodes07} Rhodes, J.~D., et al.\ 2007, \apjs, 172, 203
\bibitem[Richtler(2003)]{richtler03} Richtler, T.\ 2003, Stellar Candles for the Extragalactic Distance Scale, 635, 281
\bibitem[Riess \& Mack(2004)]{riess04} Riess, A., \& Mack, J.\ 2004, Instrument Science Report ACS 2004-006 (Baltimore: STScI), also available at http://www.stsci.edu/hst/acs/documents/isrs/isr0406.pdf
\bibitem[Robin et al.(2003)]{robin03} Robin, A.~C., Reyl{\'e}, C., Derri{\`e}re, S., \& Picaud, S.\ 2003, \aap, 409, 523
\bibitem[Sandage(1975)]{sandage75} Sandage, A.\ 1975, \apj, 202, 563
\bibitem[Schlafly \& Finkbeiner(2011)]{schlafly11} Schlafly, E.~F., \& Finkbeiner, D.~P.\ 2011, \apj, 737, 103
\bibitem[Schlegel et al.(1998)]{schlegel98} Schlegel, D.~J., Finkbeiner, D.~P., \& Davis, M.\ 1998, \apj, 500, 525
\bibitem[Schuberth et al.(2010)]{schuberth10} Schuberth, Y., Richtler, T., Hilker, M., Dirsch, B., Bassino, L.~P., Romanowsky, A.~J., \& Infante, L.\ 2010, \aap, 513, A52
\bibitem[Schulman et al.(2012)]{schulman12} Schulman, R.~D., Glebbeek, V., \& Sills, A.\ 2012, \mnras, 420, 651
\bibitem[S\'{e}rsic(1968)]{sersic68} Sersic, J.~L.\ 1968, Atlas de Galaxias Australes, Observatorio Astronomico, Cordoba
\bibitem[Sippel et al.(2012)]{sippel12} Sippel, A.~C., Hurley, J.~R., Madrid, J.~P., \& Harris, W.~E.\ 2012, \mnras, 427, 167
\bibitem[Sharina et al.(2005)]{sharina05} Sharina, M.~E., Puzia, T.~H., \& Makarov, D.~I.\ 2005, \aap, 442, 85
\bibitem[Sirianni et al.(2005)]{sirianni05} Sirianni, M., et al.\ 2005, \pasp, 117, 1049
\bibitem[Smith et al.(2013)]{smith13} Smith, R., S{\'a}nchez-Janssen, R., Fellhauer, M., et al.\ 2013, \mnras, 429, 1066
\bibitem[Spitler et al.(2006)]{spitler06} Spitler, L.~R., Larsen, S.~S., Strader, J., Brodie, J.~P., Forbes, D.~A., \& Beasley, M.~A.\ 2006, \aj, 132, 1593
\bibitem[Springel et al.(2008)]{springel08} Springel, V., Wang, J., Vogelsberger, M., et al.\ 2008, \mnras, 391, 1685
\bibitem[Taylor et al.(2010)]{taylor10} Taylor, M.~A., Puzia, T.~H., Harris, G.~L., Harris, W.~E., Kissler-Patig, M., \& Hilker, M.\ 2010, \apj, 712, 1191
\bibitem[Trager et al.(1995)]{trager95} Trager, S.~C., King, I.~R., \& Djorgovski, S.\ 1995, \aj, 109, 218
\bibitem[van den Bergh et al.(1991)]{vdB91} van den Bergh, S., Morbey, C., \& Pazder, J.\ 1991, \apj, 375, 594
\bibitem[de Vaucouleurs et al.(1991)]{RC3} de Vaucouleurs, G., de Vaucouleurs, A., Corwin, H.~G., Jr., Buta, R.~J., Paturel, G., \& Fouque, P.\ 1991, Volume 1-3, XII,~ Springer-Verlag 
\bibitem[Vesperini \& Heggie(1997)]{vesperini97} Vesperini, E., \& Heggie, D.~C.\ 1997, \mnras, 289, 898
\bibitem[Vesperini \& Zepf(2003)]{vesperini03} Vesperini, E., \& Zepf, S.~E.\ 2003, \apjl, 587, L97
\bibitem[Webb et al.(2012a)]{webb12a} Webb, J.~J., Sills, A., \& Harris, W.~E.\ 2012a, \apj, 746, 93
\bibitem[Webb et al.(2012b)]{webb12b} Webb, J.~J., Harris, W.~E., \& Sills, A.\ 2012b, \apjl, 759, L39
\bibitem[Webb et al.(2013)]{webb13} Webb, J.~J., Harris, W.~E., Sills, A., \& Hurley, J.~R.\ 2013, \apj, 764, 124
\bibitem[Wilson(1975)]{wilson75} Wilson, C.~P.\ 1975, \aj, 80, 175
\bibitem[Woodley \& G{\'o}mez(2010)]{woodley10} Woodley, K.~A., \& G{\'o}mez, M.\ 2010, PASA, 27, 379
\bibitem[Zepf et al.(1999)]{zepf99} Zepf, S.~E., Ashman, K.~M., English, J., Freeman, K.~C., \& Sharples, R.~M.\ 1999, \aj, 118, 752
\bibitem[Zibetti et al.(2009)]{zibetti09} Zibetti, S., Charlot, S., \& Rix, H.-W.\ 2009, \mnras, 400, 1181
\end{thebibliography}
\end{document}